\documentclass[utf8]{frontiersSCNS} 

\usepackage{url,lineno,microtype,subcaption}
\usepackage[onehalfspacing]{setspace}

\usepackage[english]{babel}
\usepackage{graphicx}

\def\firstAuthorLast{Matteini {et~al.}} 
\def\Authors{L. Matteini\,$^{1,2,6*}$, L. Franci\,$^{3,6}$, O. Alexandrova\,$^{2}$, C. Lacombe\,$^{2}$, S. Landi\,$^{4,6}$, P. Hellinger\,$^{5}$, E. Papini\,$^{4,6}$ and A. Verdini\,$^{4,6}$}
\def\Address{$^{1}$Department of Physics, Imperial College London, London SW7 2AZ, UK\\
$^{2}$LESIA, Observatoire de Paris, Universit\'e PSL, CNRS, Sorbonne Universit\'e, Univ. Paris Diderot, Sorbonne Paris Cit\'e, France\\
$^{3}$School of Physics and Astronomy, Queen Mary University of London, UK \\
$^{4}$Dipartimento di Fisica e Astronomia, Universit\'a di Firenze, Italy\\
$^{5}$Astronomical Institute, CAS, Prague, Czech Rep.\\
$^{6}$ INAF, Osservatorio Astrofisico di Arcetri, Firenze, Italy}
\def\corrAuthor{Imperial College London, South Kensington Campus, London SW7 2AZ, UK}

\def\corrEmail{l.matteini@imperial.ac.uk}

\begin{document}
\onecolumn
\firstpage{1}

\title[Properties of sub-ion turbulence]{Magnetic field turbulence in the solar wind at sub-ion scales: in situ observations and numerical simulations}

\author[\firstAuthorLast ]{\Authors} 
\address{} 
\correspondance{} 

\extraAuth{}

\maketitle

\begin{abstract}

We investigate the transition of the solar wind turbulent cascade from MHD to sub-ion range by means of a detail comparison between in situ observations and hybrid numerical simulations. In particular we focus on the properties of the magnetic field and its component anisotropy in Cluster measurements and hybrid 2D simulations.
First, we address the angular distribution of wave-vectors in the kinetic range between ion and electron scales by studying the variance anisotropy of the magnetic field components. When taking into account the single-direction sampling performed by spacecraft in the solar wind, the main properties of the fluctuations observed in situ are also recovered in our numerical description. This result confirms that solar wind turbulence in the sub-ion range is characterized by a quasi-2D gyrotropic distribution of k-vectors around the mean field.

We then consider the magnetic compressibility associated with the turbulent cascade and its evolution from large-MHD to sub-ion scales. The ratio of field-aligned to perpendicular fluctuations, typically low in the MHD inertial range, increases significantly when crossing ion scales and its value in the sub-ion range is a function of the total plasma beta only, as expected from theoretical predictions, with higher magnetic compressibility for higher beta. Moreover, we observe that this increase has a gradual trend from low to high beta values in the in situ data; this behaviour is well captured by the numerical simulations. The level of magnetic field compressibility that is observed in situ and in the simulations is in fairly good agreement with theoretical predictions, especially at high beta, suggesting that in the kinetic range explored the turbulence is supported by low-frequency and highly-oblique fluctuations in pressure balance, like kinetic Alfv\'en waves or other slowly evolving coherent structures. The resulting scaling properties as a function of the plasma beta and the main differences between numerical and theoretical expectations and in situ observations are also discussed.
\end{abstract}

\twocolumn

\section{Introduction}
The solar wind constitutes a unique laboratory for plasma turbulence \citep[][]{Bruno_Carbone_2013}.
In the last decade increasing interest has raised towards the small-scale behaviour of the turbulent cascade, i.e. beyond the breakdown of the fluid/MHD description that takes place at ion scales. 
Spacecraft observations of solar wind and near-Earth plasmas provide unique measurements of the turbulent fluctuations at scales comparable and smaller than the typical particle scales, the Larmor radius $\rho$ (see Appendix for definition of physical quantities used) and the inertial length $d$ \citep[e.g.][]{Alexandrova_al_2009, Sahraoui_al_2010, Alexandrova_al_2012, Chen_al_2013}.
However, the physical processes governing the energy cascade at kinetic scales and those responsible for its final dissipation are not well understood yet.

What is well established is that in the transition from MHD to kinetic regime, plasma turbulence modifies its characteristics.
Observational and numerical studies over the last few years have highlighted the main differences between large and small scale properties of solar wind fluctuations \citep[e.g.,][]{Chen_2016, Cerri_al_2019}.
The magnetic field spectrum typically steepens when approaching ion scales, leading at sub-ion scales (between ion and electron typical scales) to a power law with spectral index close to $-2.8$ \citep{Alexandrova_al_2009, Alexandrova_al_2012, Kiyani_al_2009, Chen_al_2010, Sahraoui_al_2013}, steeper than Kolmogorov -5/3,  but also than the theoretical prediction $-7/3$ from  EMHD \citep{Biskamp_al_1996} and KAW/whistler turbulence \citep{Schekochihin_al_2009, Boldyrev_al_2013}. 
The origin of such a spectral slope is still unknown and it has been proposed that it could be related to intermittency corrections \citep{Boldyrev_Perez_2012, Landi_al_2019}, magnetic reconnection \citep{Loureiro_Boldyrev_2017, Mallet_al_2017, Cerri_al_2018}, Landau damping \citep{Howes_al_2008, Schreiner_Saur_2017} and/or the role of the non-linearity parameter \citep{Passot_Sulem_2015, sulem_al_2016}.

The change in the magnetic field spectrum is accompanied by a rapid decrease in the power of ion velocity fluctuations \citep{Safrankova_al_2013, Stawarz_al_2016} and the onset of the non-ideal terms in the Ohm's law which governs the electric field associated to the turbulent fluctuations; as a consequence the electric field spectrum becomes shallower at sub-ion scales \citep{Franci_al_2015b, Matteini_al_2017}. In this framework the electric current (mostly carried by electrons) plays a major role, coupling directly with the magnetic field in the cascade and likely affecting the energy cascade rate via the Hall term \citep{Hellinger_al_2018, Papini_al_2018}. All these properties depend further on the plasma beta ($\beta=8\pi nk_BT/B^2$), which controls, among other things, the scale at which the magnetic field spectrum breaks \citep{Chen_al_2014b, Franci_al_2016}.

One of the most significant differences with respect to the turbulent regime observed at large scales however is the role of compressive effects. 
While in the inertial range fluctuations show a low level of both plasma and magnetic field compressibility, hence can be reasonably well described by incompressible MHD, at sub-ion scales density and magnetic field intensity fluctuations become significant and comparable to transverse ones \citep{Alexandrova_al_2008,  Sahraoui_al_2010, Salem_al_2012, Kiyani_al_2013, Chen_al_2012, Perrone_al_2017}, in agreement with simulations \citep{Franci_al_2015a, Cerri_al_2017}. 
It is believed that this is related to a change in the properties of the turbulent fluctuations, which become intrinsically compressive at small scales.
It is then by studying in detail their properties that it is possible to shed light on the nature of the fluctuations which support the cascade at kinetic scales \citep{Chen_al_2013b, Pitna_al_2019, Groselj_al_2019, Alexandrova_al_2020}.

Another important aspect of solar wind turbulence is its spectral anisotropy \citep{Horbury_al_2008,Wicks_al_2010,Chen_al_2010,Roberts_al_2017}. Studies about the shape of turbulent eddies, both at MHD \citep{Chen_al_2012b, Verdini_al_2018, Verdini_al_2019} and kinetic scales \citep{Wang_al_2020}, reveal the presence of a  3D anisotropy in the structures when described in terms of a local frame. On the other hand, when the analysis is made in a global frame (without tracking the local orientation of the structures), the 3D anisotropy is not captured, and the k-vectors of the fluctuations show a statistical quasi-2D distribution around the magnetic field \citep{Matthaeus_al_1990, Dasso_al_2005, Osman_Horbury_2006}. In this work we address this latter aspect and we investigate the distribution of the k-vectors with respect to the ambient magnetic field at kinetic scales by using the magnetic field variance anisotropy (i.e. the ratio of magnetic field fluctuations in different components). \cite{Saur_Bieber_1999} have shown that, also in single spacecraft observations, is possible to characterize the 3D k-vector distribution by using variance anisotropy. When the sampling occurs only along a preferential direction, like in typical solar wind observations, their model predicts various possible kinds of variance anisotropy as a function of the underlying k-spectrum. In particular,  assuming a quasi-2D gyrotropic distribution of k-vectors (axisymmetric with respect to the magnetic field), the ratio of the power in the two perpendicular magnetic field components is directly related to the local slope of the spectrum - which is assumed to have the same form for all components and a slope independent of the scale within a given regime. Since both quantities, spectral slope and perpendicular power ratio, can be easily measured in situ, the Saur \& Bieber model constitutes an useful and simple tool to investigate underlying spectral anisotropies.
Despite the model was originally developed for MHD scale fluctuations, it basically corresponds to a geometrical description built on the divergence-less condition for $\boldsymbol{B}$, so it can be applied to any kind of regimes, including the low-frequency turbulence expected at sub-ion scales \citep{Turner_al_2011}. In the work of \cite{Lacombe_al_2017}, we investigated the k-vector distribution at sub-ion scales using the technique by \cite{Saur_Bieber_1999}. Based on the comparison with the predictions, we concluded that the distribution of the k-vectors in the sub-ion range of solar wind turbulence is consistent with a quasi-2D gyrotropic spectrum, then approaching a more isotropic shape when reaching electron scales \citep{Lacombe_al_2017}. However, such an application has not been benchmarked by kinetic numerical studies yet.

The aim of this work is then to focus on the spectral anisotropy properties and magnetic compressibility at small scales, by exploiting the detailed comparison of in situ observations and high-resolution kinetic numerical simulations. The paper is organised as follow: in Sec. \ref{method} we introduce the spacecraft and numerical dataset used and in Sec. \ref{spectra} we describe their spectral  properties. In Sec. \ref{sec_anis} we discuss the spectral anisotropy at sub-ion scales and test, for the first time, the Saur and Bieber model in numerical kinetic simulations; in Sec. \ref{sec_comp} we address properties of the magnetic compressibility and its dependence on the plasma beta. Finally, in Sec. \ref{conclusion} we discuss our conclusions and the implications of our finding for the interpretation of solar wind observations and simulations.

\begin{figure*}[t]
\includegraphics[width=18cm]{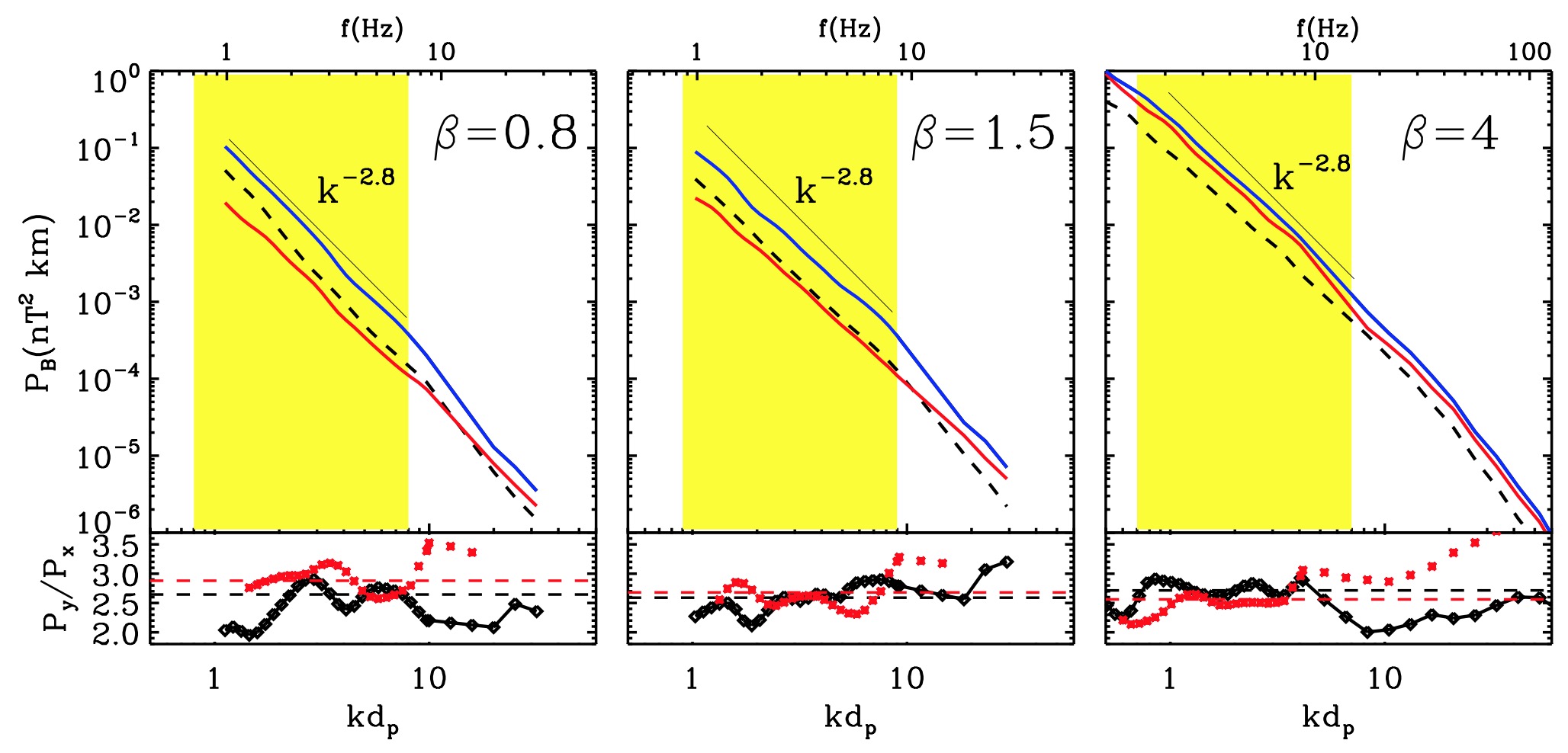}
\caption{Cluster STAFF spectra for different intervals with $\beta=0.8, 1.5, 4$ from left to right. Colors encode the magnetic field components $B_x$ (black dashed), $B_y$ (blue),  and $B_z$ (red). The region highlighted in yellow corresponds to the the sub-ion range investigated in this study. Bottom panels show the ratio of the power in the perpendicular components $P_y(k)/P_x(k)$ (black diamonds) and the value $\gamma$ of the local slope of the total spectrum $P(k)\sim k^{-\gamma}$ (red stars). The average values of $P_y(k)/P_x(k)$ and $\gamma$ in the sub-ion range are also shown as horizontal dashed lines. \label{fig1}}
\end{figure*}

\begin{figure*}
\includegraphics[width=18cm]{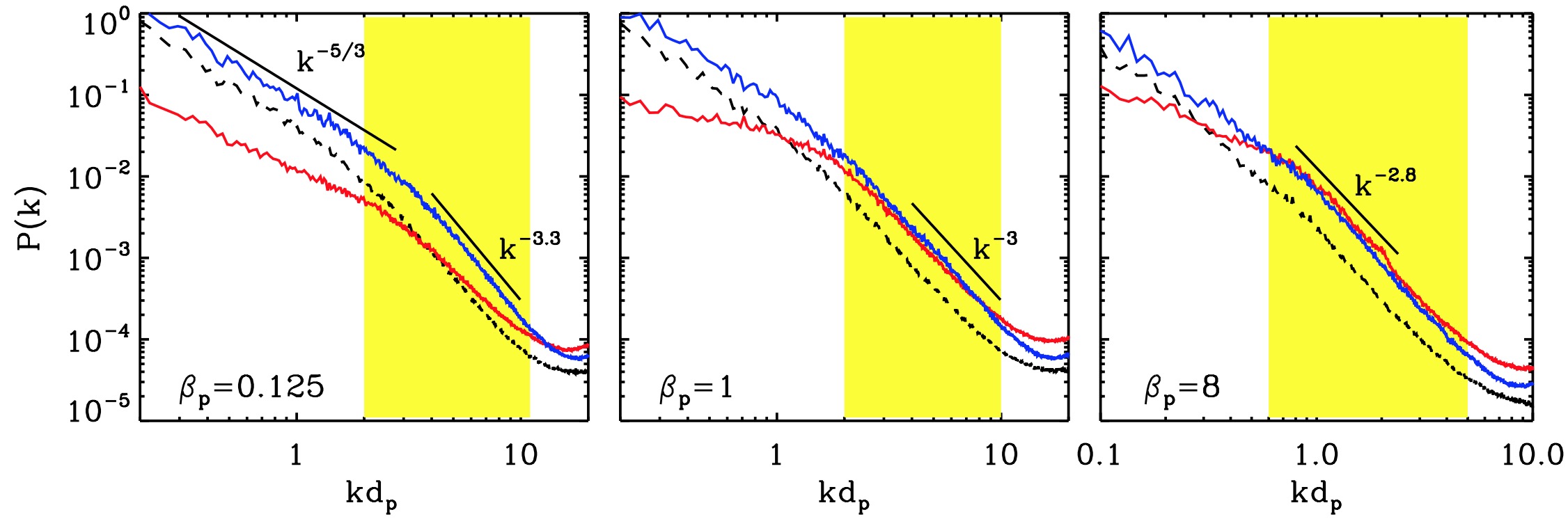}
\caption{Magnetic field spectra from hybrid simulations for different beta regimes ($\beta_p=0.125,1,10$ and $\beta_e=\beta_p$). Components are encoded as in Figure~\ref{fig1} and the coloured region indicates the sub-ion range that can be directly compared with the analogous region in the observations.
\label{fig2}}
\end{figure*}

\section{Data and simulations}\label{method}
In this study we compare properties of magnetic fluctuations measured in situ by the Cluster spacecraft with numerical results obtained by means of 2D hybrid particle-in-cell (PIC) simulations.

\subsection{Cluster STAFF spectra}
For our analysis we use the dataset discussed by \citet{Alexandrova_al_2012}, when Cluster was in the free solar wind, i.e., not magnetically connected to the Earth's bow shock. Details have been described also in \cite{Lacombe_al_2017} and we recall here the main aspects. Magnetic field fluctuations are measured by the STAFF (Spatio-Temporal Analysis of Field Fluctuation) instrument, composed by a wave form unit (SC) and a Spectral Analizer (SA). Power spectra are computed on board in a magnetic field aligned system of coordinates (MFA), based on the 4s magnetic field measured by the FGM (Fluxgate Magnetometer) experiment.
A selection of 112 spectra has been performed, retaining in each spectrum only measurements above 3 times the noise level in every direction $x, y$ and $z$ \citep[see Appendix in][]{Lacombe_al_2017}. Each sample is a 10 minute average of 150 individual 4s spectral measurements. 
This provides spectra above 1Hz up to typically 20-100Hz, depending on the amplitude of the fluctuations in each interval.
When converted into physical length scales, assuming the Taylor hypothesis ($k=2\pi f/V_{sw}$), this leads to signals that cover the range between $\sim2d_p$ and $\sim0.5d_e$ (where $d_p$ and $d_e$ are the proton and electron inertial lengths respectively), enabling then a good description of the sub-ion regime from proton to electron scales.

The reference frame adopted (MFA) is such that $B_z$ is the component aligned with the mean magnetic field $B_0$ (relative to the 4s interval during which an individual spectrum is calculated); $B_x$ is the component orthogonal to $B_z$ in the plane containing both the solar wind velocity $V_{sw}$ and the mean magnetic field $B_0$, and $B_y$ is the third orthogonal component.
Note that a selection criterium is imposed on the angle $\theta_{BV}$, the angle between the local 4s magnetic field and the flow velocity, i.e., that $\theta_{BV}$ is large enough to avoid a connection with the Earth bow shock during the sampled interval; $\theta_{BV}$ in the dataset has an average value of $\sim80$ degrees.
This implies that for each spectrum, the mean magnetic field makes a big angle with respect to the sampling direction; moreover, we have checked that $\theta_{BV}$ does not vary significantly during the 10 minutes over which spectra are averaged. 

As a consequence, this procedure selects intervals in which Cluster observed highly oblique k-vectors and, to a good approximation, the component $B_x$ corresponds also to the sampling direction (radial) and is orthogonal to $B_0$; $B_y$ corresponds to the other perpendicular component and $B_z$ is identified as the compressive component $B_\|$.
As already discussed in \cite{Lacombe_al_2017}, although the total trace power measured in situ is an invariant observable, the fact that the sampling occurs only in a preferred direction introduces a relative weight between 
$B_x$ and $B_y$ that is measurement dependent \citep{Saur_Bieber_1999}.
To take this into account, we have employed an analogous approach in the analysis of the simulations data, as described in the next section.

\subsection{Hybrid 2-D numerical simulations}
In situ observations are directly compared with numerical simulations performed with the hybrid-PIC code CAMELIA \citep{Matthews_1994, Franci_al_2018b}.
The hybrid model captures well the transition from fluid to kinetic regime around ion scales. Moreover, it reproduces successfully many of the main properties of solar wind turbulence observed by spacecraft at sub-ion scales \citep{Franci_al_2015a, Franci_al_2015b}. It is then a suitable tool to investigate the turbulent regime probed by STAFF/Cluster data. 
We use here 2D simulations -computationally more affordable than 3D- in order to explore the parameter space observed in situ; in particular we focus on the effects associated to variations in the proton and electron plasma beta $\beta_p$ and $\beta_e$. \cite{Franci_al_2016} have shown that 2D hybrid simulations are able to capture the ion-break scale behaviour in different beta regimes observed in solar wind turbulence \citep{Chen_al_2014b}. We then exploit the good matching between the simulations and in situ observations to characterise further the properties of kinetic plasma turbulence in the sub-ion regime.
On the other hand, 3D hybrid simulations \citep{Franci_al_2018} have confirmed the solidity of the reduced 2D results and the good agreement with in situ observations.

In order to make a direct comparison with sub-ion spectra measured by Cluster, we have adopted a similar approach in the computation of spectra in the simulations. This means that numerical spectra are computed along the $x$ direction only, to mimic the radial sampling occurring in the solar wind. This is obtained by integrating along $y$ the Fourier spectrum $P(k_x,y)$ of each $i$ magnetic field component: 
\begin{equation}
P_i(k_x)=\int{P_i(k_x,y)dy}
\end{equation}
Therefore, also in the simulation $B_x$ corresponds to the sampling direction, orthogonal to the out-of-plane magnetic field $B_z$, and $B_y$ is the most energetic fluctuating component, being orthogonal to both $B_0$ and $k=k_x$. With this approach and within the observational conditions previously described, we can perform a direct comparison of simulations and in situ data.

The numerical dataset used was originally presented in \cite{Franci_al_2016} and is available online. It is constituted by a set of different $2048^2$ 2-D simulations of decaying turbulence with different beta conditions covering the range of variations observed in situ, and $\beta_p=\beta_e$; runs are initiated with random perpendicular Alfv\'enic fluctuations with vanishing cross-helicity and equipartition in magnetic and kinetic energies. Spectra are computed at the maximum of the turbulent activity.

\section{In situ data analysis and simulation results}\label{spectra}

Figure~\ref{fig1} shows three examples of Cluster spectra (2003/02/18 04:45-04:55; 2004/02/22 05:40-05:50; 2004/01/22 04:40-04:50), where frequencies have been converted into k-vectors and normalised to $d_p$ (original sampling frequencies are also shown for reference). Observations cover ion and electron scales, with a transition accompanied by a slope change around $kd_p\sim10$ . In this work, we focus on the sub-ion regime highlighted in yellow in the panels, where electron physics effects can be neglected (at least for spectral properties) and  a well-defined slope close to $-2.8$ can be observed \citep{Alexandrova_al_2012}. The three cases, corresponding to different total beta $\beta$ regimes $[0.5, 1.5, 4]$, show a similar qualitative behaviour: as expected, the spectrum $P_y$ of the perpendicular $B_y$ component (blue) is always the most energetic. The power in the other perpendicular component $P_x$ (black dashed) is always slightly smaller, however, its ratio with $P_y$ is roughly independent of beta and close to the local spectral slope (bottom panels); this is related to the 3-D distribution of k-vectors \citep{Lacombe_al_2017} and will be discussed more in detail in Sec.\ref{sec_anis}. 

\begin{figure}
\includegraphics[width=8.cm]{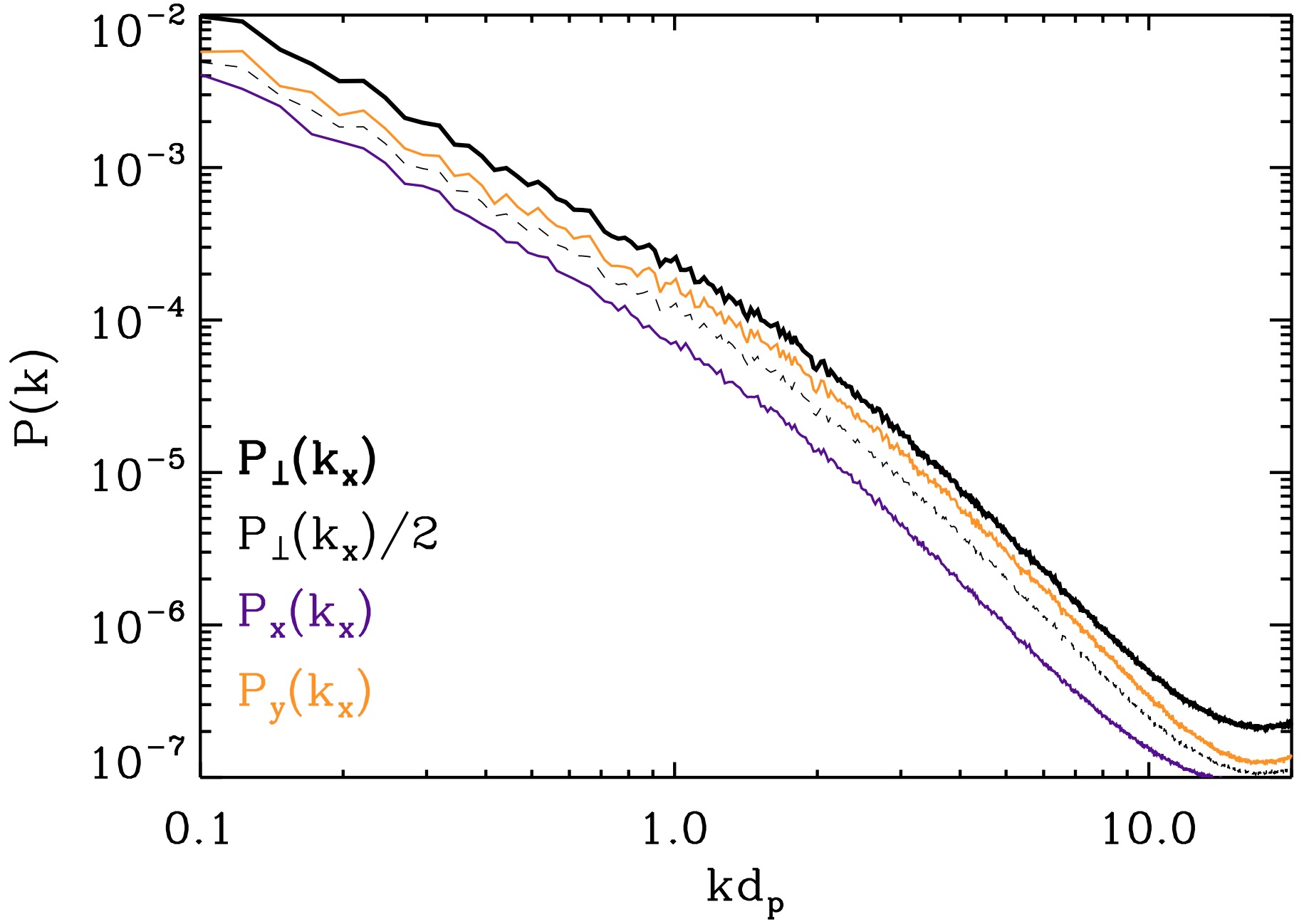}
\caption{Reduced spectra of the fluctuations of the magnetic field components $B_y$ and $B_x$ defined with respect to a fixed sampling direction $k_x$ for a simulation with $\beta_p=0.5$. The thick solid black line corresponds to the total perpendicular power $P_\perp(k_x)$; the dashed line shows $P_\perp(k_x)/2$, also corresponding to the average power in any perpendicular magnetic field component in the axisymmetric case. \label{fig3}}  
\end{figure}

On the other hand, the power $P_z$ of the field aligned component $B_z$ (red) is typically less energetic than $P_y$, however, its weight is highly variable with beta: $P_z$ is smaller than $P_x$ for $\beta<1$, comparable to $P_x$ for  $\beta\sim1$, and larger the $P_x$ for $\beta>1$. This obviously results in a variable magnetic compressibility associated to the fluctuations and its functional dependence on beta is the subject of Sec.\ref{sec_comp}

Figure~\ref{fig2} shows an analogous selection from numerical simulations; note that in the simulations $\beta_e=\beta_p$. In this case the regime reproduced in the simulation box includes the MHD inertial range and its transition to a sub-ion cascade at smaller scales. The yellow area highlights the region of the spectra $-$ roughly a decade between $kd_p\sim1$ and $kd_p\sim10$ $-$  that can be directly compared with the in situ data. In this region, the qualitative behaviour of the spectra is similar to Figure~\ref{fig1}: $B_y$ (blu) is always dominant, $B_x$ (black) contributes for a constant fraction of it and is roughly the same at all betas, while $B_z$ (red) varies significantly in the panels and becomes comparable to $B_y$ for large betas.
This confirms that our method of computing spectra in the simulations mimicking satellite observations really captures the main aspects of in situ measurements and can then be exploited to investigate further the properties of the turbulent cascade.

\section{Spectral anisotropy}\label{sec_anis}

\subsection{Perpendicular components ratio}
\cite{Saur_Bieber_1999} have investigated how different types of k-vectors distributions can generate a variable anisotropy in the observed magnetic field components, due to sampling effects. In the case of a gyrotropic 2D distribution of k-vectors, the ratio $P_{y}/P_{x}$ is expected to coincide with the local slope $\gamma$ of the spectrum $P(k)\sim k^{-\gamma}$.
This applies well to solar wind observations in the physical range of interest here, as it can be appreciated in Fig.~\ref{fig1}, where the ratio $P_{y}/P_{x}$, shown in the bottom panels, is close to the spectral slope observed - typically in the range [-2.5,-3] - and appears roughly independent of the plasma beta. Interestingly, at smaller scales, when the magnetic spectrum steepens as approaching electron scales \citep{Alexandrova_al_2009}, this is not associated to an increase in the perpendicular power ratio $P_{y}/P_{x}$ (which on the contrary has a slight decrease); this doesn't correspond to the expectation for a quasi-2D spectrum according to the model and in fact \cite{Lacombe_al_2017} has interpreted this signature as the result of a more isotropic distribution of k-vectors close to electron scales.

To validate further this observational conclusion, we verify here the applicability of the Saur and Bieber model to sub-ion scale turbulence. In the simulations the spectrum is two-dimensional by construction and, consistent with the axisymmetric initial conditions imposed in the $x$-$y$ plane, it is also gyrotropic with respect to the out-of-plane magnetic field $B_z$.

First, it is instructive to discuss spectra shown in Figure~\ref{fig3}. These are power spectra of the perpendicular components $B_x$ (purple) and $B_y$ (orange) as a function of $k_x$, assuming then a fixed direction of sampling. As expected, $P_y(k_x)>P_x(k_x)$; on the other hand, their sum $P_\perp(k_x)$ (solid black line) is statistically equivalent to the axisymmetric spectrum $P_\perp(k)=P_y(k)+P_x(k)$. The difference is that when calculating the axisymmetric spectrum $P_\perp(k)$ all perpendicular magnetic field directions have equal weight and one can assume that statistically $P_y(k)\sim P_x(k)$; as a consequence the power associated to any individual perpendicular component corresponds to half of the total perpendicular power $P_\perp(k)/2\sim P_\perp(k_x)/2$ (thin dashed black line). It is interesting to note that when sampling along a fixed direction ($x$), as it happens with spacecraft in the solar wind, none of the two measured spectra $P_y(k_x)$ and $P_x(k_x)$ is really representative of the power $P_\perp(k)/2$ of the gyrotropic description; instead the component along the sampling ($B_x$) is significantly reduced due to the solenoidal $\boldsymbol{\nabla}\cdot\boldsymbol{B}=0$ condition, while the orthogonal ($B_y$) is amplified, in order to maintain the same total power $P_\perp(k)$.
This means that in solar wind spectra like in Fig.~\ref{fig1}, neither $P_x$ nor $P_y$ are individually representative of the average power in a perpendicular B component: the individual measurements of $P_x$ or $P_y$ cannot be directly associated to it, but only their sum. 

\begin{figure}
\includegraphics[width=8.cm]{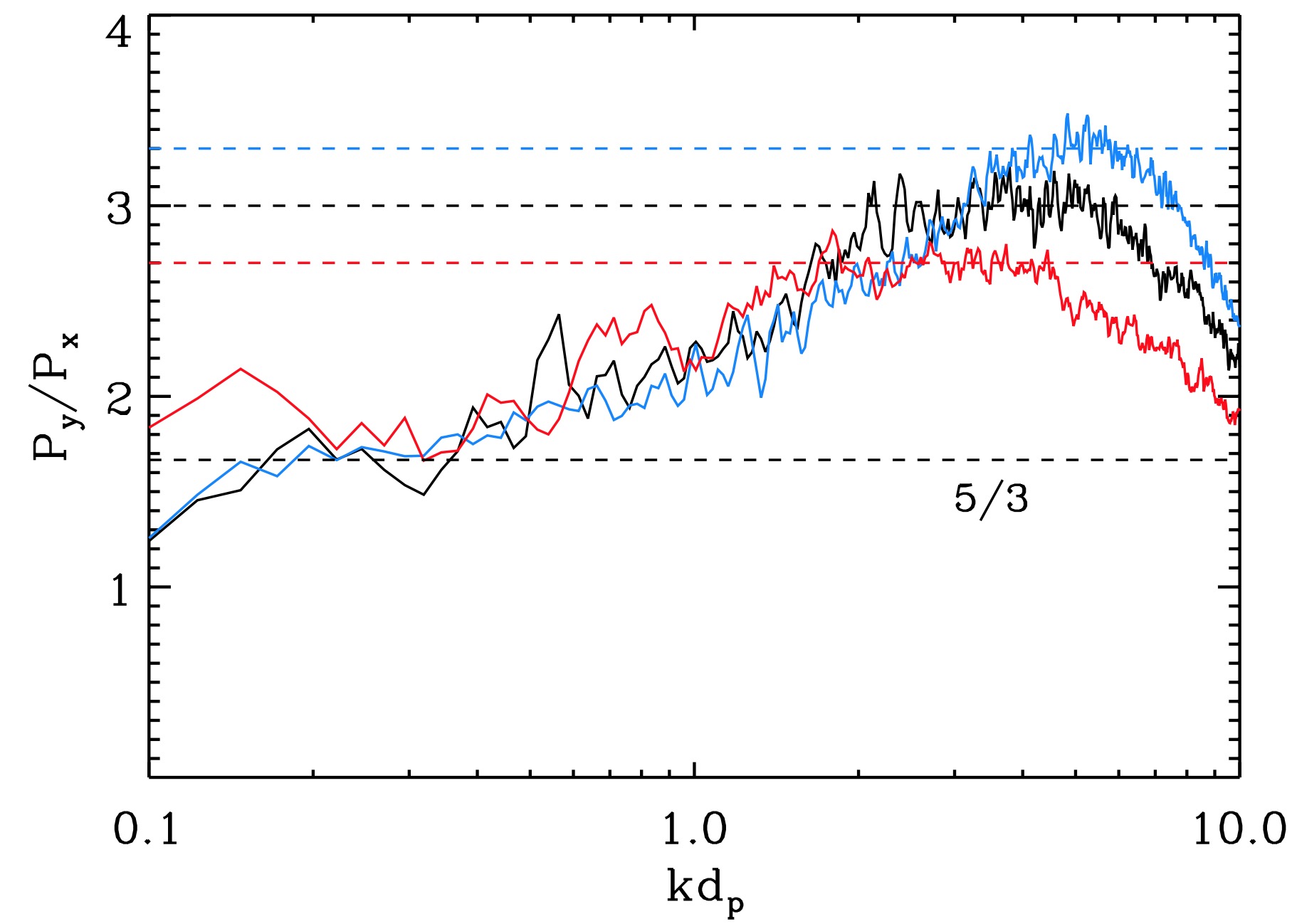}
\caption{Hybrid simulations: spectra of the ratio of the perpendicular magnetic components $P_y(k_x)/P_x(k_x)$ and corresponding to the local spectral slope. Different colors encode different $\beta_p$: 0.125 (cyan), 1 (black) and 8 (red). The horizontal dashed lines show reference spectral slopes observed in the simulations at $kd_p<1$ (-5/3) and at sub-ion scales $kd_p>1$. \label{fig4}}  
\end{figure}

Bearing this in mind, Figure~\ref{fig4} shows the ratio of the power in the perpendicular components for the three simulations shown in Figure~\ref{fig2}. The $P_y/P_x$ ratio well captures the transition from MHD to a steeper spectrum at smaller scales; in all cases the ratio, close to 5/3 at large scales, starts increasing in the vicinity of ion scales and reach a maximum in the sub-ion regime, where the ratio saturates close to $\sim3$, in good agreement with the local spectral slope observed in the kinetic range, which is typically close to $-3$. At larger $k$ the ratio then decreases due to the noise. In the framework of the Saur and Bieber model for spectral anisotropy this indicates a quasi-2D gyrotropic spectrum of the fluctuations, which corresponds well to the spectrum developed in these simulations. This confirms that the model is valid also at sub-ion scales, and reinforces the finding of \cite{Lacombe_al_2017}, where is found that solar wind spectra at kinetic scales are described well by a quasi-2D gyrotropic distribution.

\subsection{Beta dependence} 
\begin{figure}
\includegraphics[width=8.5cm]{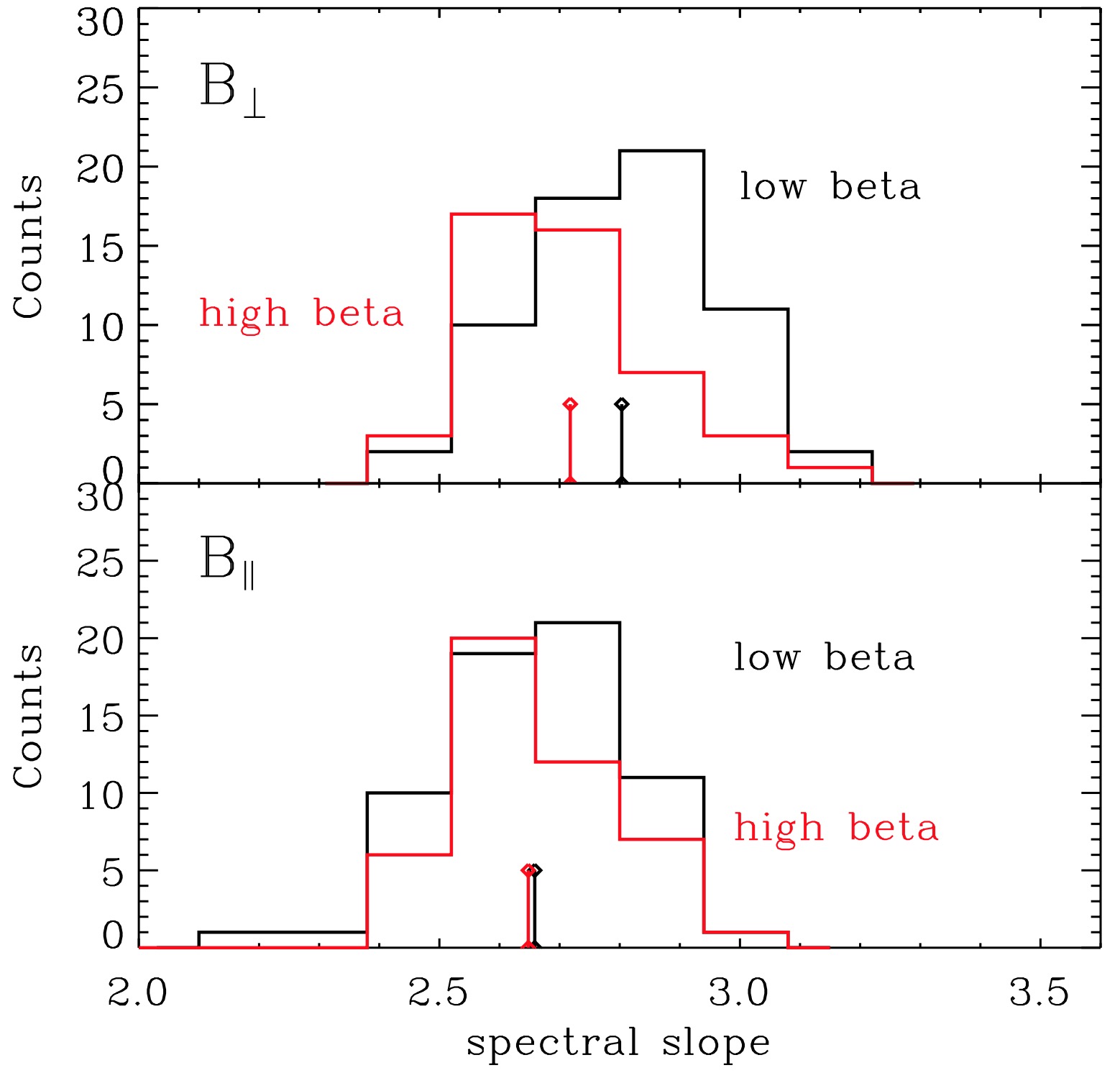}
\caption{Spectral slope measured for different beta conditions in Cluster data; (black) $\beta<2$ and (red) $\beta>2$. Top panel refers to the spectrum of the perpendicular magnetic component $B_\perp$ and the bottom panel to $B_\|$. The short vertical lines ending with a diamonds indicate average values of the histograms.\label{fig5}}  
\end{figure}

There is another interesting indication suggested by Figure~\ref{fig4}, namely the fact that the $P_y/P_x$ ratio in the sub-ion range seems to depend on beta: consistent with this, the sub-ion slope in Figure~\ref{fig2} is slightly steeper for small $\beta_p$ and shallower for larger $\beta_p$. This behaviour is already discussed in \cite{Franci_al_2016} and is found in all simulations for the spectrum of the transverse fluctuations $B_\perp$; conversely, the spectrum of the parallel component $B_\|$ is almost independent of $\beta_p$ 
\cite[see Figure 4 in][]{Franci_al_2016}. We have then looked for a similar trend also in the in situ data. Figure~\ref{fig5} shows the histogram of the spectral slopes in the kinetic range for $B_\perp$ (top) and $B_\|$ (bottom), for larger (red) and smaller (black) total beta. Spectral slopes are calculated between $2<kd_p<8$ for $\beta_p<1$ and  between $2<k\rho_p<8$ for $\beta_p>1$, where a quite well-defined power law scaling is observed. They are then separated in two groups defined by the total beta $\beta<2$ and  $\beta>2$. The mean of each histogram is indicated by the small vertical line ended with a diamond.
For the parallel component (bottom panel), the distribution of the slopes is similar for both beta regimes and centred around a value of approximately $-2.65\pm0.15$; this is in good agreement with the simulations. For the dominant perpendicular component (top panel), we observe average values consistent with previous studies based on the total power $\delta B^2=\delta B_\|^2+\delta B_\perp^2$ of the fluctuations \citep{Alexandrova_al_2009, Alexandrova_al_2012, Sahraoui_al_2013, Chen_al_2013}. However, in the lower beta case (black), some slightly steeper slopes are observed for $B_\perp$ with respect to the high beta case, with an average of $-2.8\pm0.15$ with respect to $-2.7\pm0.15$.
This behaviour agrees qualitatively with the simulations; however a more detailed investigation is needed to fully identify the role of $\beta_p$ on the sub-ion spectral slope and is beyond the scope of the present study.

What can be however pointed out is that a consequence of the behaviour in Figure~\ref{fig5} is that while at high beta, $\delta B_\|$ and $\delta B_\perp$ are observed to have almost the same scaling, so that their ratio remains approximately constant in the sub-ion range, at lower $\beta$ their slightly different scaling is expected to result in a slow increase of the $\delta B_\|/\delta B_\perp$ ratio between ion and electron scales. These properties are related to the evolution of the magnetic compressibility of the fluctuations in the sub-ion range, which is main focus of the next section.

\begin{figure}
\includegraphics[width=8.cm]{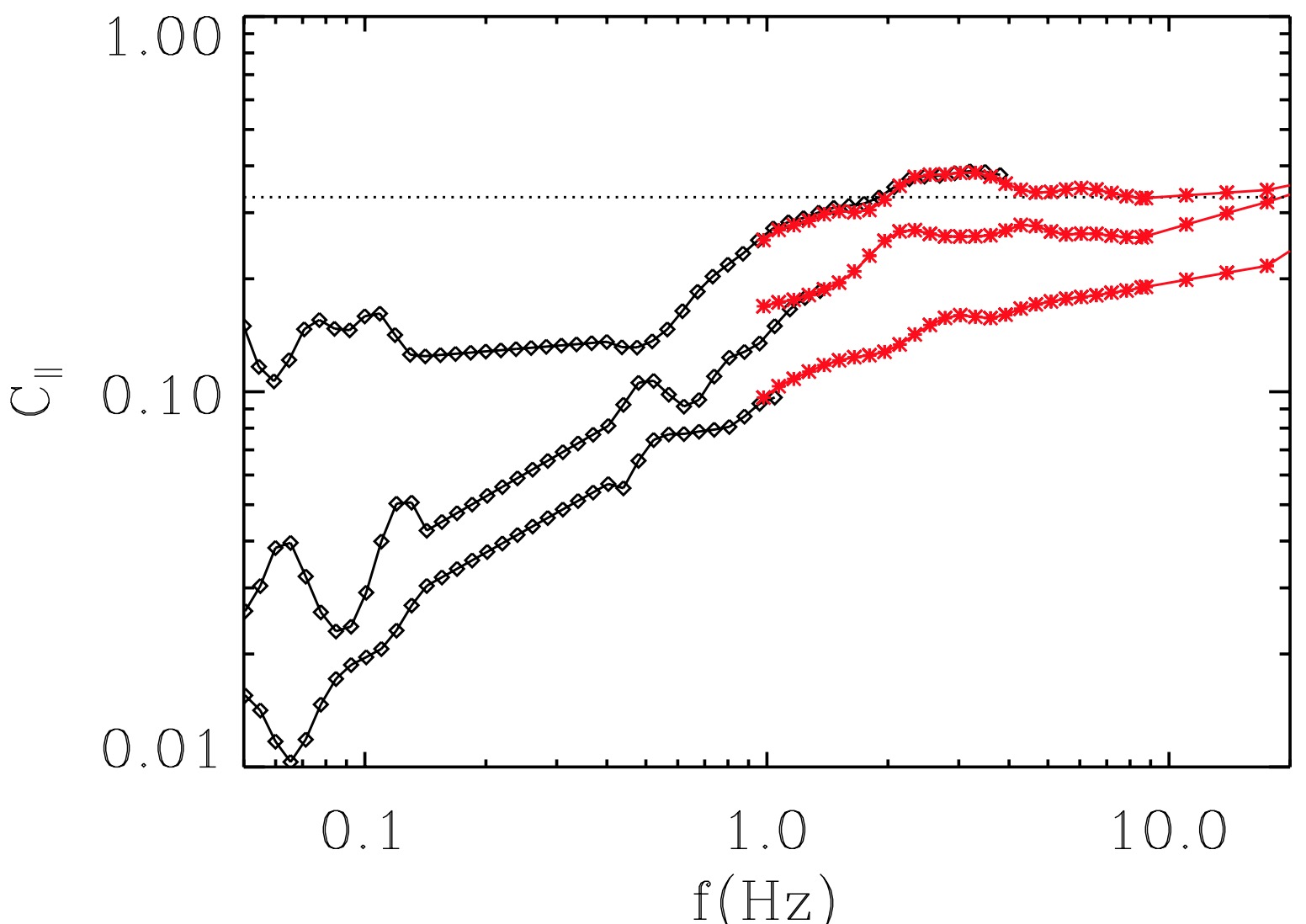}
\caption{Examples of Cluster FGM (black) and STAFF (red) spectra of magnetic compressibility $C_\|$ as a function of the frequency measured in spacecraft frame.\label{fig6}}  
\end{figure}

\section{Magnetic compressibility}\label{sec_comp} 
We now investigate the role of the third magnetic field component $B_z$, which is aligned with the local (at 4s) magnetic field $B_0$. In particular, we focus on the 
the magnetic compressibility $C_\|=\delta B_{\|}^2/\delta B^2$, where $\delta B^2=\delta B_\|^2+\delta B_\perp^2$ and its implication for the nature of the cascade at these scales. Note that in this case, the measurement of $B_z$ is not affected by the sampling direction (provided that this is orthogonal to $B_0$ to a good approximation) and since we use the total perpendicular power $P_\perp$, the caution discussed in Sec.~\ref{sec_anis} is not needed here.

Figure~\ref{fig6} shows $C_\|$ for three intervals of different total $\beta$=1,3,4 ($\beta_p=0.3, 1.4, 2.5$) as measured from STAFF (red). For these three cases we also show the spectrum of the magnetic field compressibility as measured at lower frequencies (corresponding to physical scales larger than $d_p$) by the fluxgate magnetometer onboard Cluster (FGM, black). Note that FGM spectra are linearly interpolated between $0.14$ and $0.4$Hz to remove artefacts due to spacecraft spin ($0.25$ Hz). There is a good matching between the two independent measurements at $f\sim1$Hz and where data points from both instruments are available for a more extended range there is also a quite satisfactory overlap between them.
The overall behaviour agrees well with the expected picture: at large scale, in the MHD inertial range, the level of compressibility is lower, typically $C_\|\lesssim0.1$ \citep[e.g.,][]{Horbury_Balogh_2001, Smith_al_2006b} and starts to increase as approaching ion scales \citep{Hamilton_al_2008, Alexandrova_al_2008, Salem_al_2012, Kiyani_al_2013}, reaching sometimes variance isotropy (indicated by the dashed horizontal line) in the sub-ion range, where the compressibility seems to saturate. As already shown by \cite{Lacombe_al_2017}, the level of magnetic compressibility developed at small scales is larger for high beta than for small beta. 
Since we focus on the behaviour at sub-ion scales, in the following we restrict our analysis to STAFF measurements only.

\begin{figure}
\includegraphics[width=9.cm]{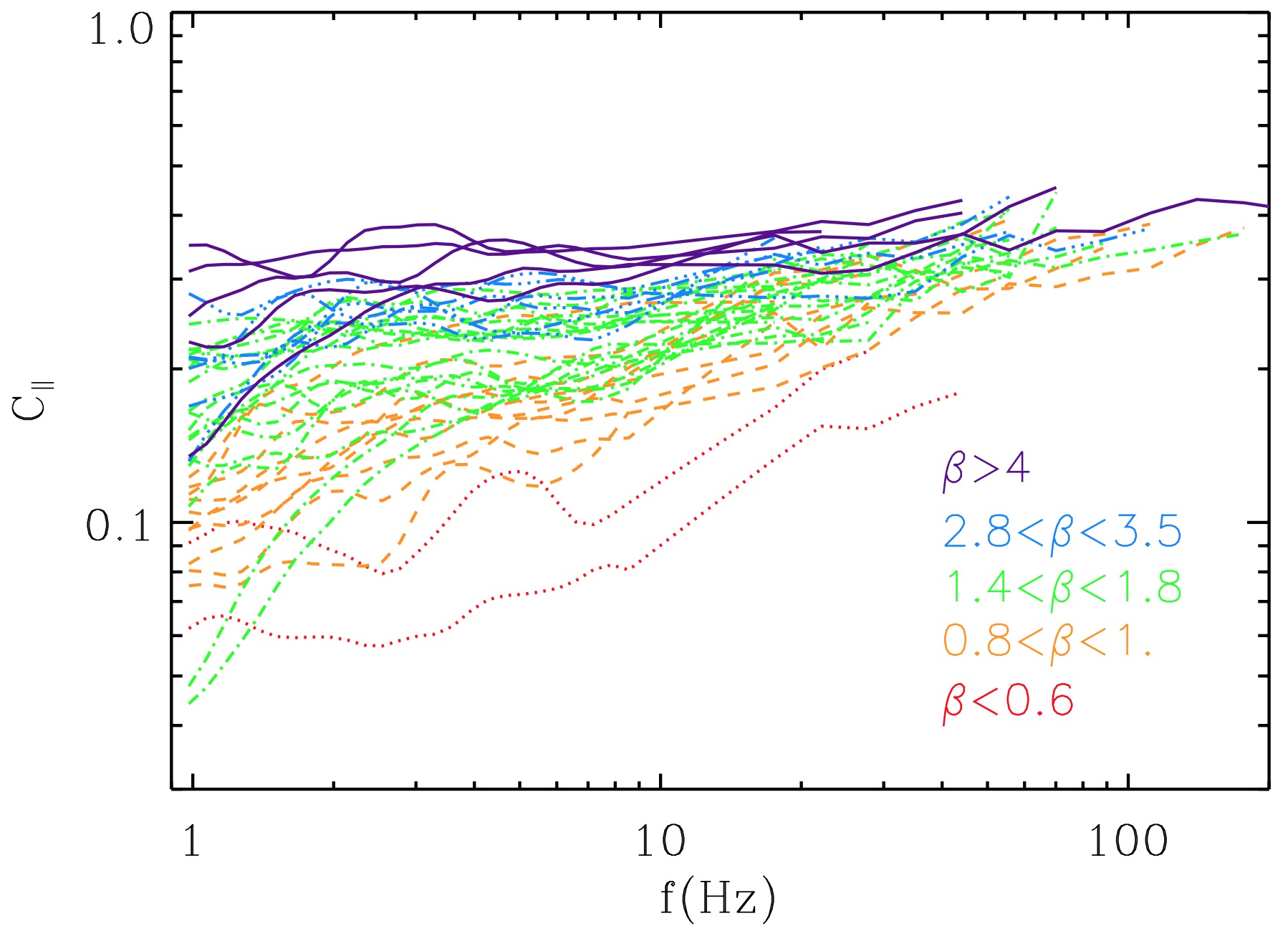}
\caption{STAFF spectra of magnetic compressibility $C_\|$ as a function of the frequency measured in spacecraft frame. Different colours and lines identify different groups of intervals with given $\beta$.\label{fig7}}
\end{figure}

To highlight further the $\beta$-dependence of the magnetic compressibility,  Figure~\ref{fig7} shows $C_\|$  for a selection of spectra with different $\beta$, increasing from red to purple. There is a continuous transition from lower to higher magnetic compressibility as a function of beta, in agreement with linear theory expectations \citep[e.g.][]{Podesta_TenBarge_2012}.
Moreover, at high beta it seems that the fluctuations reach an asymptotic $\delta B_{\|}^2/\delta B^2$ ratio, leading to an extended plateau in the spectrum, while at the lowest beta a plateau cannot be clearly identified.
We now want to identify more in detail what process and length scale control the level of $C_\|$ and in solar wind data. 

\subsection{Beta dependence and theoretical predictions}

\begin{figure}
\includegraphics[width=8.5cm]{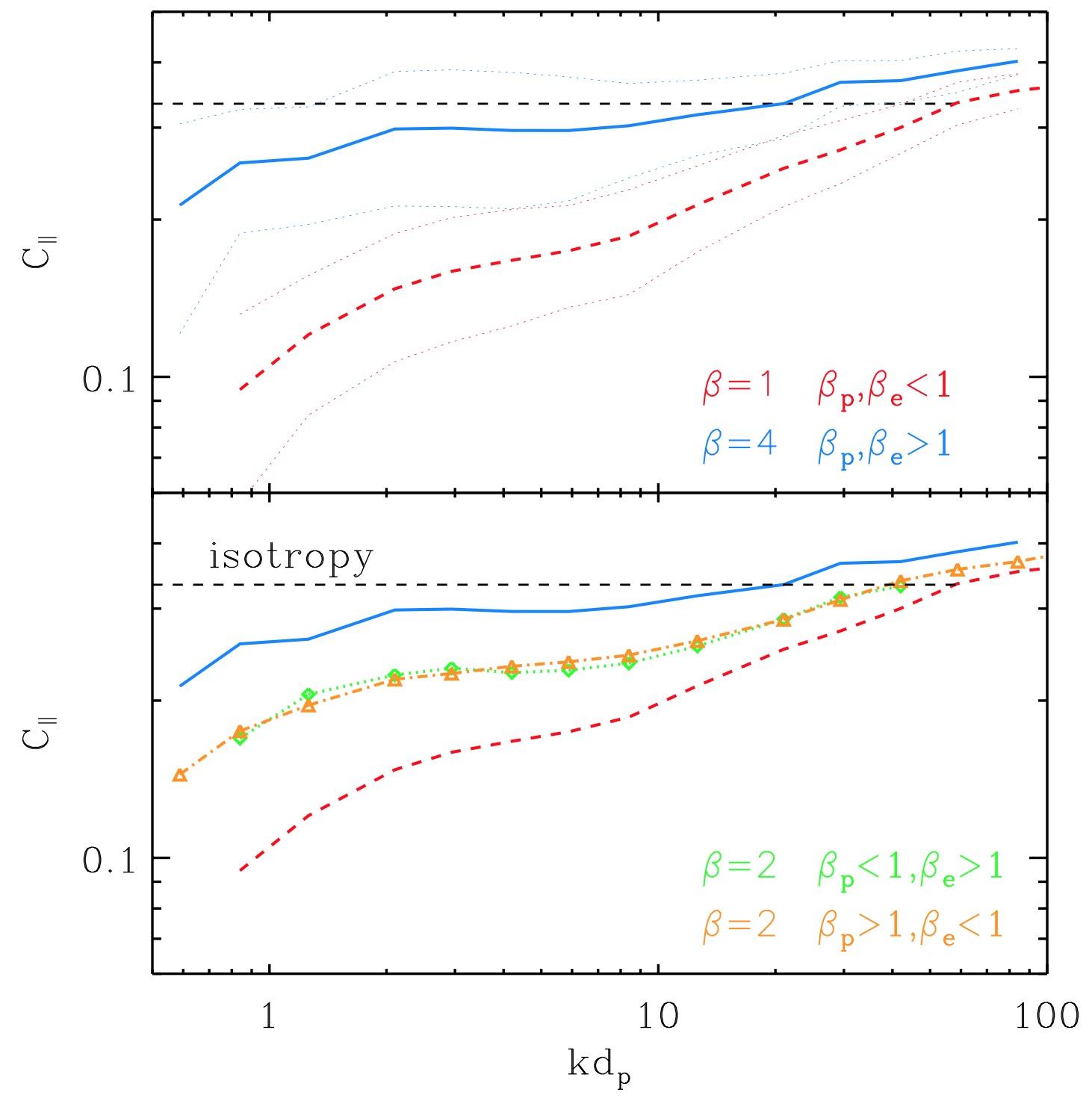}
\caption{Cluster average spectra of magnetic compressibility for intervals with $\beta_e$ and $\beta_p<1$ (red) and $\beta_e$ and $\beta_p>1$ (blue); in the bottom panel, intermediate values with similar average $\beta$ but with $\beta_p<1$, $\beta_e>1$, and $\beta_p>1$, $\beta_e<1$ are also shown in green and orange respectively. Thin dotted lines in the upper panel show the one-sigma dispersion of the data.}
\label{fig8} 
\end{figure}

First it is useful to go again from frequency to k-vector spectra:
in Figure~\ref{fig8} frequencies are converted into k-vectors and normalised with respect to the proton inertial length $d_p$. 

We first identify two big categories such that both proton and electron betas are small, i.e. $\beta_p<1$ and $\beta_e<1$, or both large, i.e. $\beta_p>1$ and $\beta_e>1$.  We obtain an average total beta $\beta\sim1$ in the former, and $\beta\sim4$ in the latter. The average spectrum of magnetic compressibility for each of the two families is shown in the top panel of Figure~\ref{fig8} as a function of  $kd_p$; the thin dotted lines identify the standard deviation around the averages. 
In the high beta case (solid blue) the compressibility reaches a plateau after $kd_p=1$ and saturates at an average level which is very close to isotropy (same power in $P_{x}, P_{y}$ and $P_{z}$), while in the low beta case (dashed red) $C_\|$ remains smaller and there is not a clear plateau at $kd_p>1$.

The remaining spectra are further separated in two other families: the first with $\beta_p<1$ and $\beta_e>1$, the second $\beta_p>1$ and $\beta_e<1$. In this case the average total betas are very similar, $\beta\sim1.9$ ($\beta_p\sim0.75$) and $\beta\sim2.0$ ($\beta_p\sim1.5$), respectively, and fall in between the other two groups (small and large $\beta$). Consistent with this, the average spectrum of these two families, shown in orange and green in the bottom panel, has a level of compressibility at sub-ion scales that is intermediate with respect to the other two curves. Moreover, they almost precisely fall on top of each other. All this suggests that not only the total plasma beta is a good parameter for ordering the level of compressibility generated at sub-ion scales, but also this level is roughly independent on the individual weights of $\beta_p$ and $\beta_e$, being their sum $\beta=\beta_p+\beta_e$ the only relevant parameter.

This observational finding is in very good agreement with the expectation from the relation below:
\begin{equation}\label{eq_balance}
C_\|=\frac{\beta_p/2(1+T_e/T_p)}{1+\beta_p(1+T_e/T_p)}=\frac{\beta/2}{1+\beta}
\end{equation}
where $T_e$ and $T_p$ are the electron and proton temperatures.

Eq. (\ref{eq_balance}) can be derived \citep{Schekochihin_al_2009, Boldyrev_al_2013} under the assumption of low-frequency magnetic structures in pressure balance at scales where the ion velocity becomes negligible compared to the electron one, or equivalently the Hall term $\boldsymbol{J} \times \boldsymbol{B}$ becomes dominant over the ideal MHD term $-\boldsymbol{U}\times \boldsymbol{B}$. A special case is the regime of kinetic Alfv\'en waves (KAW), however eq. (\ref{eq_balance}), which does not depend explicitly on $k$ and thus on a specific dispersion relation, can be seen as a more general condition for highly oblique fluctuations in the sub-ion range \citep[e.g. ion-scale Alfv\'enic vortices,][]{Jovanovic_al_2020}, under the assumptions described above \cite[see e.g. Appendix C2 of][]{Schekochihin_al_2009}. 

\begin{figure*}
\includegraphics[width=17.5cm]{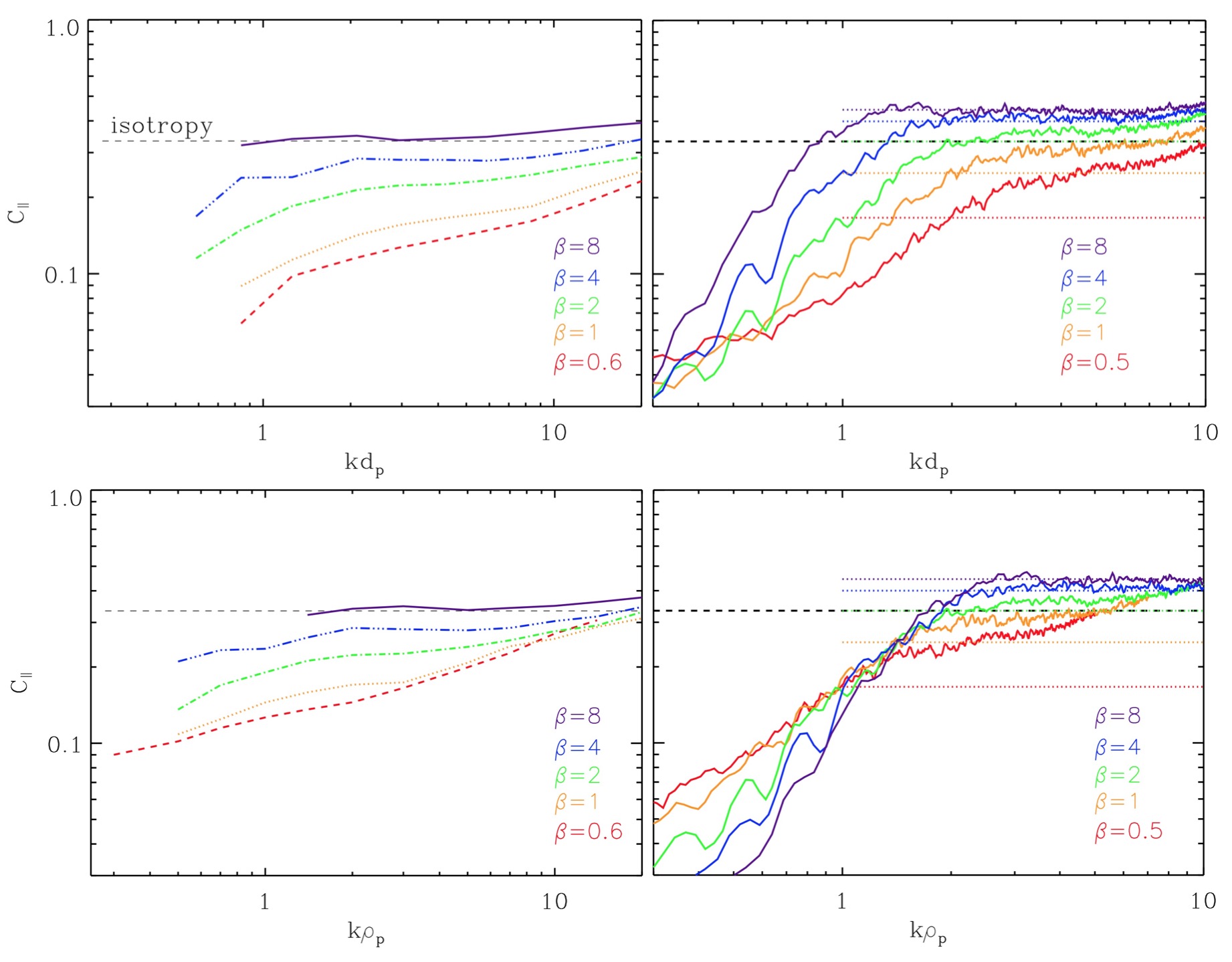}
\caption{Top panels: (Left) Cluster spectra of magnetic compressibility for intervals binned on different $\beta$, encoded in different styles and colours. Only cases with $\beta_p\sim\beta_e$ have been retained. The horizontal dashed line denotes energy equipartition between components (i.e. isotropy). (Right) Spectra of magnetic compressibility for simulations with different $\beta_p=\beta_e$, shown with same style as left panel. The increase of $C_\|$ for $kd_i\gtrsim8$ is due to numerical noise. Bottom: Same as top panels, but with k-vectors normalised with respect to the ion gyro-radius $\rho_p$. Horizontal dotted lines, coloured according to their $\beta$, are the theoretical prediction of $C_\|$ from Eq.(\ref{eq_balance}).
\label{fig9}}
\end{figure*}

\subsection{Comparison with simulations}

To improve our analysis we focus more in detail on the Cluster observations and compare them with numerical results. Note that as in the simulations of \cite{Franci_al_2016} it is only considered the case $\beta_p=\beta_e$, we have made a selection of solar wind spectra with similar properties ($\beta_p\sim\beta_e\sim\beta/2$). These have then been divided in 5 sub-groups as a function of $\beta$ and averaged to obtain a mean $C_\|$ profile for each  $\beta$-family. The selection results in 7, 13, 23, 9 and 1 spectra for $\beta=0.6,1,2,4,8$ respectively (only 1 spectrum fulfils the condition for high enough beta). Simulations with approximatively the same $\beta_p$ (and $\beta$) are considered for a direct comparison.
In the following analysis we want to identify the physical scale associated to the changes in the properties of the fluctuations and its possible connection to either the ion Larmor radius $\rho_p$ or the inertial length $d_p$, as they are related by: $\rho_p=\sqrt{\beta_p}d_p$. 

The results of this comparison are shown in Figure~\ref{fig9}, where scales are normalised to both $d_p$ (top) and $\rho_p$ (bottom). Left panels show spectra from in situ data and right panels results from simulations, where the colors encode the same range of $\beta$. 
Qualitatively, the global trend seen in the simulations matches well that of the observations. First, the level of magnetic compressibility reached at sub-ion scales increases monotonically with $\beta$, as expected. Second, we can identify a plateau phase beyond ion scales whose extension is gradually reduced as $\beta$ decreases; for the smallest betas the plateau disappears and is replaced by an almost monotonic increase of $C_\|$ all along the sub-ion range - though with a shallower slope compared to that of the transition from the MHD range. 

This seems to suggest a different behaviour of the turbulent fluctuations populating the sub-ion cascade as a function of the beta.
To investigate further this aspect, horizontal dotted lines in the right panels of Figure~\ref{fig9} show the theoretical prediction for the asymptotic level of $C_\|$ between ion and electron scales predicted by Eq.(\ref{eq_balance}), with same color scale.
For simulations at large $\beta$, when a plateau is clearly observed, the level of magnetic compressibility also agrees well with the one predicted by the theory. In the low beta case there is a larger discrepancy and the observed level of magnetic compressibility is larger than the constant level predicted by Eq. (\ref{eq_balance}). The different behaviour of the compressibility in low- and high-beta regimes found in our simulations, together with the larger discrepancy with respect to the theoretical predictions observed at low beta, are also consistent with results from previous numerical studies \citep[e.g.][]{Cerri_al_2016, Cerri_al_2017, Groselj_al_2017}.

The situation is somewhat different when comparing predictions to the in situ data; in this case there is a slight difference between the KAW level and the observed one, and this is persistent at all $\beta$. 
In particular, at high beta it is apparent that while Eq.~(\ref{eq_balance}) predicts a compressibility that goes beyond 1/3 (for $\beta\to\infty$ we have $C_\|=0.5$, so $\delta B_\|=\delta B_\perp$), a condition well recovered in the simulations, in Cluster data $C_\|$ does not go beyond component isotropy ($\delta B_\|=\delta B_\perp/2$, thus  $C_\|=1/3$). However, due to the low statistics in the data (just 1 spectrum has $\beta\gtrsim8$) it's hard to draw a firm conclusion here.  

\begin{figure*}
\includegraphics[width=17.5cm]{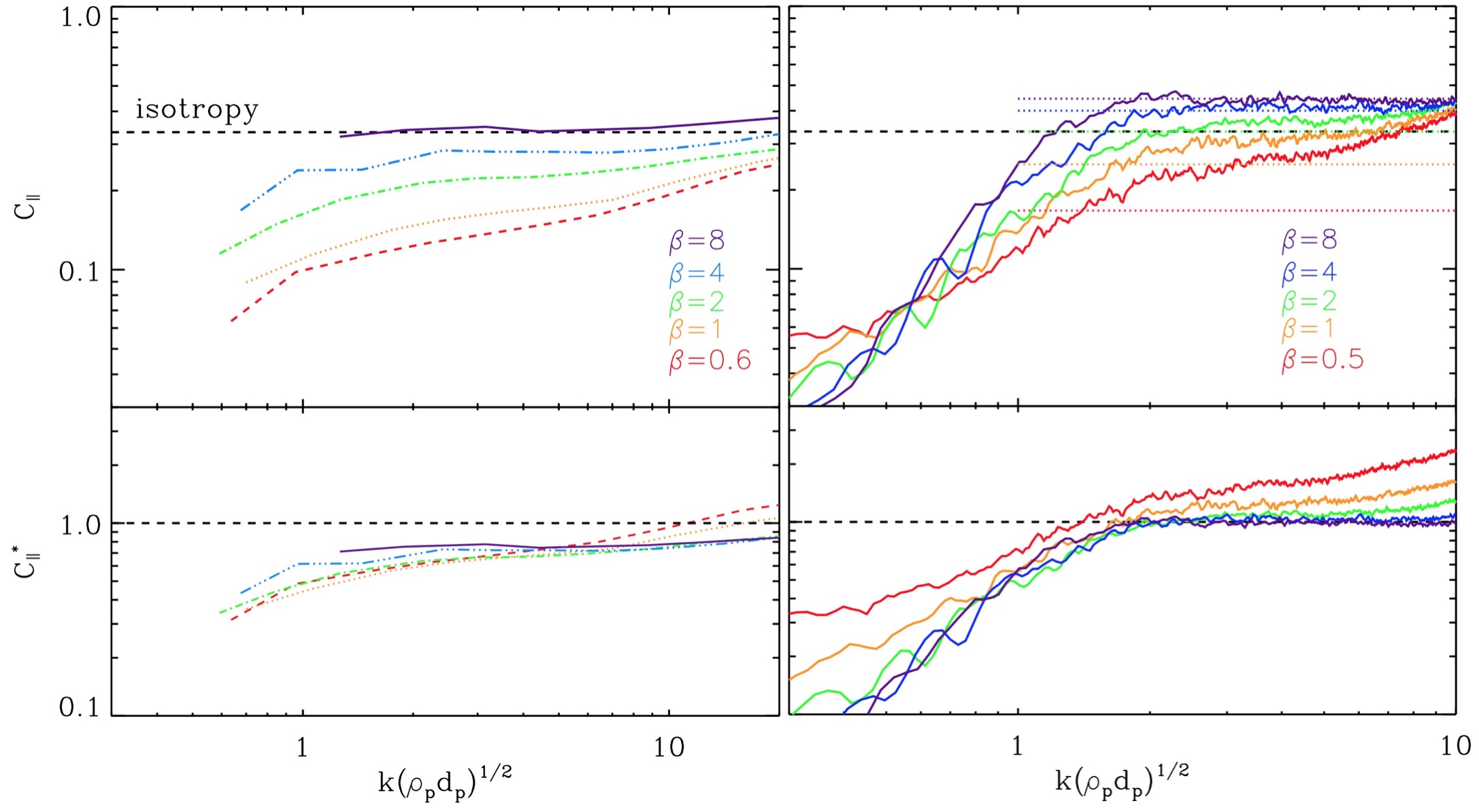}
\caption{Same spectra as in Figure~\ref{fig9}, but with k-vectors normalised to the mixed scale $\sqrt{d_p\rho_p}$; in the bottom panels $C_\|$ is normalised to the theoretical prediction by Eq.\ref{eq_balance}.
\label{fig10}}
\end{figure*}

Interestingly, from Figure~\ref{fig9} it seems that nor $d_p$ nor $\rho_p$ are able to fully capture and order the change in the spectrum of the magnetic compressibility for different betas;  the saturation/plateau phase for low $\beta$ spectra results more shifted towards high k-vectors compared to the high $\beta$ ones when normalising to $d_p$, while the vice-versa is observed when normalising to $\rho_p$. This suggests that the behaviour can be better captured by an intermediate scale between the two. For this reason, in Figure~\ref{fig10} we have normalised spectra to a mixed scale $\sqrt{d_p\rho_p}$. Note that such a scale, proportional to $d_p\beta_p^{1/4}$, was found to describe well the behaviour of ion-break scale in magnetic field spectra in the range $\beta_p\sim1$ by \cite{Franci_al_2016}, and, although not shown, to describes the variation of the break of the parallel magnetic field spectrum at all betas; this then motivated our choice. When such a mixed scale is used (top right panel), all cases follow the same trend: they grow until they reach $k\sqrt{d_p\rho_p}\sim2$ and then start flattening, the saturation level depending on the beta. In situ observations (top left panel) seem to follow the same trend, confirming that such an intermediate scale is a good candidate for controlling the variation of the magnetic compressibility spectrum at ion scales. 

It is then reasonable to use such a k-vector normalization to better evaluate the agreement with Eq.~(\ref{eq_balance}). In the bottom panels of the same figure $C_\|^*$ spectra are then normalised to the theoretical prediction for $C_\|$. 
In simulations, as already pointed out, cases with $\beta>1$ display a good agreement with the sub-ion compressibility level predicted by the theory; as a consequence, when normalised to $\sqrt{d_p\rho_p}$ all spectra collapse on top of each other all along ion and sub-ion scales. A worse agreement is observed at $\beta\le1$ when simulations display a slightly higher compressibility level than predicted.
Quite differently, the ratio between the in situ observations and the theoretical $C_\|$ is always below 1 and around 0.7-0.8 for all $\beta$ groups in the sub-ion range \cite[see also Figure 10 of][]{Lacombe_al_2017}. This behaviour is consistent with the results of \cite{Pitna_al_2019} based on observations from the Wind spacecraft, who find on average $C_\|\sim0.9$ - without making a distinction among beta regimes - and with most of the data displaying a slightly smaller magnetic compressibility than the prediction. Our study confirms this scenario and suggests that the same trend is followed for all spectra, almost independently of the plasma beta.
A ratio smaller than 1 and close to $~\sim0.75$ is also consistent with similar observational results of the plasma compressibility and based on the ratio between density and perpendicular magnetic fluctuations predicted by linear theory \citep{Chen_al_2013, Pitna_al_2019}. This was interpreted by \cite{Chen_al_2013} as a consequence of the non-linear behaviour of the solar wind fluctuations in the sub-ion range, in agreement with simulations of strong KAW-turbulence \citep{Boldyrev_al_2013}. On the other hand, for the magnetic compressibility, our fully non-linear simulations of sub-ion turbulence do not recover the same effect seen in situ, as $C_\|^*\gtrsim1$.
Other reasons could explain such a discrepancy, e.g., the effect of some electron Landau damping on the fluctuations observed in situ \citep{Howes_al_2011, Passot_Sulem_2015, Schreiner_Saur_2017} and not captured by the hybrid model. In order to answer these questions, a more detailed study of the polarisation properties of the fluctuations in our simulations is in preparation.

Finally, note that the increase in $C_\|$ observed at higher $k$ in the in situ data could be related to a further change in the properties of the fluctuations as they approach electron scales; as discussed in \cite{Lacombe_al_2017} this also coincides with a change in the estimated spectral anisotropy. For example, \cite{Chen_Boldyrev_2017} have suggested that the increase in the magnetic compressibility beyond the sub-ion range could be related to electron inertia corrections to Eq.\ref{eq_balance}. This effect is then not captured by the hybrid model and we cannot compare any more the observations with the simulations in this range. It is however interesting to note that while the further increase of compressibility at electron scales is predicted for $\beta_e\lesssim1$ \citep{Chen_Boldyrev_2017, Passot_al_2017}, in the intervals measured by Cluster it seems to be observed for all beta ranges for $kd_p\gtrsim10$ ($kd_e\gtrsim1/4$). Moreover, it's also interesting to note that spectra for all betas reach isotropy at roughly $k\rho_p\sim20$, corresponding on average to $k\rho_e\sim0.5$.

\section{Conclusion}\label{conclusion}
In summary, we have discussed properties of magnetic field spectra of turbulent fluctuations in the sub-ion regime and their main dependence on the plasma beta. We have carried out a detailed comparison between in situ Cluster magnetic field observations in the frequency range $f$(Hz)$=[1,200]$, corresponding to scales typically between ${d_p<l<d_e}$, and high-resolution 2D hybrid simulations. 

First we investigated the spectral anisotropy of magnetic fluctuations at sub-ion scales. Our simulations confirm that the model of \cite{Saur_Bieber_1999}, originally developed for MHD range fluctuations, is valid also at kinetic scales; by applying the model to the numerical spectra obtained mimicking the sampling along a fixed direction made by spacecraft, we were able to successfully capture original spectral properties as well as their variation with $\beta$. This then reinforces the finding of \cite{Lacombe_al_2017} who applied the Saur and Bieber model to kinetic-scale observations for the first time and concluded that fluctuations of the solar wind spectrum in the sub-ion range are quasi-2D and gyrotropic. Moreover, we have shown that the component anisotropy measured in situ - leading to an apparent non-gyrotropic spectrum from an original gyrotropic one \citep[see also][]{Turner_al_2011}  - is a direct consequence of the solenoidal condition of the magnetic field and the sampling procedure. This is not an effect related to the Doppler-shift of k-vectors swept through the spacecraft by the fast plasma flow and in fact, we were able to reproduce it in simulations just imposing a fixed sampling direction.

Note that our result about the global 2D-symmetry of the k-vectors around the magnetic field is not inconsistent with studies addressing the local shape of the eddies and suggesting the presence of a 3D anisotropy \citep[e.g.][]{Chen_al_2012b, Verdini_Grappin_2015, Verdini_al_2018, Verdini_al_2019, Wang_al_2020}. In our approach we do not consider the specific orientation of the turbulent structures in the plane perpendicular to B, and it is reasonable to expect that their local 3D anisotropy is then lost. In other words, despite the possible presence of a 3D anisotropy of the turbulent eddies their k-vectors can be oriented isotropically around B, leading then - in a frame like the one used here - to the 2D spectrum found in the Cluster observations. 
This does not exclude that some aspects of the 3D anisotropy could be still captured using also a global approach, however, our study suggests that in this case one has to take carefully into account the effects of the apparent component anisotropy introduced by the sampling \cite[][see also Figure~\ref{fig3} in this work]{Saur_Bieber_1999}.

For the magnetic compressibility $C_\|$, we have confirmed that it has a strong dependence on the plasma beta \citep[e.g.,][]{Alexandrova_al_2008, TenBarge_al_2012, Lacombe_al_2017}. In particular we have shown that  in Cluster observations $C_\|$ depends on the total beta $\beta$ only (Fig.~\ref{fig8}), as expected for low-frequency pressure-balanced fluctuations at highly oblique propagation (e.g., KAW). In the $\beta$ range explored we find a good qualitative agreement between the trend observed in the data and in the simulations. The compressibility is observed to increase as a function of $\beta$, leading to a plateau at sub-ion scales for high betas and in good agreement with the prediction by Eq.~(\ref{eq_balance}). At low beta, a full developed plateau is not observed beyond ion scales and the compressibility continue to slowly increase along sub-ion scales, both in observations and simulations \citep[see also][]{Groselj_al_2019}. 
There is, however, a difference in the asymptotic level of compressibility reached at high $\beta$ in data and our simulations; in the former, fluctuations seem not to exceed component isotropy ($C_\|=1/3$), while in the latter they approach $C_\|=0.5$, which is the limiting value predicted by Eq. (\ref{eq_balance}). This aspect deserves to be explored in future studies, extending the range of $\beta$ explored, to then establish if the asymptotic condition observed in simulations and predicted by the theory, which implies same power in the parallel component as in the sum of the perpendicular ones, can be also observed in situ for high enough $\beta$ intervals.
As a consequence of the behaviour just described, there is a different quantitative agreement of the magnetic compressibility observed in situ and in simulations, with the theoretical prediction by Eq. (\ref{eq_balance}). In simulations there is very good matching with the predicted level at higher beta, but an excess of $C_\|$ at low beta; this effect was already observed in \cite{Cerri_al_2017} and is confirmed here on a large range of $\beta$.
On the other hand, in solar wind observations, the ratio is always lower than 1 (smaller compressibility than predicted by the theory), and close to $\sim0.75$ for all $\beta$, in agreement with similar studies on the plasma compressibility \citep{Chen_al_2013, Pitna_al_2019}.

Our analysis also suggests that the increase in the compressibility at ion scales is controlled by an intermediate scale between the Larmor radius $\rho_p$ and the proton inertial length $d_p$ (Fig.~\ref{fig9}). For simulations this was already anticipated in \cite{Franci_al_2016}, and we could identify it as related to $\sqrt{d_p\rho_p}$, thus proportional to $d_p\beta_p^{1/4}$ (Fig.~\ref{fig10}). Such a scaling with $\beta_p$ also corresponds to the scaling observed for the spectral ion-break in the range $\beta_p\sim1$.  However, it is worth to highlight that both observations \citep{Chen_al_2014b} and our simulations \citep{Franci_al_2016} show that the spectral ion-break scale follows the largest of $\rho_p$ and $d_p$ depending on the beta, so that the correction term proportional to $d_p\beta_p^{1/4}$ identified in \cite{Franci_al_2016} is important only around $\beta_p\sim1$. On the other hand, 
the present study indicates that a scale proportional to $\sqrt{d_p\rho_p}$ orders well the spectra of compressibility at all betas, for both in situ data and simulations, suggesting that such a mixed scale controls the transition in the nature of the fluctuations from MHD to sub-ion range \cite[see also the monotonic scaling with $\beta_p$ of the ion-break in the parallel magnetic field spectrum shown in Fig.4 of][]{Franci_al_2016}. This may imply that the two changes of regime - the steepening of the magnetic spectrum and the increase in the compressibility - can occur at different scales for more extreme $\beta$ values. In particular, we expect the spectral break to occur at a larger scale with respect to the plateau in the compressibility when $\beta_p\gg1$ or $\beta_p\ll1$, as in these cases $\sqrt{d_p\rho_p}$ is always smaller than the largest between $\rho_p$ and $d_p$.
A more detailed analysis on this aspect will be the subject of a future study, as well as the possible implications of this behaviour for fluctuations in the inner Heliosphere, where the plasma beta is typically lower than at 1AU, which can be observed by the Parker Solar Probe and Solar Obiter.

\section*{Appendix: Symbol definitions and  
normalized units}
The subscripts $\perp$ and $\parallel$ refer to the direction with respect to the
ambient magnetic field
$\boldsymbol{B}_0$ and $p$ and $e$ denote respectively protons and electrons.
All equations are expressed in the c.g.s. unity system.
$n$ and $T$ denote the number density and the temperature of
a species (we assume also $n_p=n_e=n$).
$\beta_{e,p}=8\pi n k_B T_{e,p}/B_0^2$ are the
electron and proton betas, and $\beta=\beta_p+\beta_e$ is the total plasma beta; here $k_B$ is the
Boltzmann constant. 
For each species of mass $m$ and charge $q$, the inertial length $d$ is defined a $c/\omega_{p}$, where $\omega_p=(4\pi n q^2/m)^{1/2}$ is the plasma frequency, and 
the Larmor radius $\rho$ is defined as $v_{th}/{\Omega_{c}}$ where $v_{th}$ it the thermal speed of each species and 
$\Omega_c=q_p B_o/m c$ is the cyclotron frequency. $V_{sw}$ is the solar wind speed and $f$ the frequency of the fluctuations measured by the spacecraft; $k$ denotes the module of the wave vector $\boldsymbol{k}$.

\section*{Conflict of Interest Statement}
The authors declare that the research was conducted in the absence of any commercial or financial relationships that could be construed as a potential conflict of interest.

\section*{Author Contributions}
LM anf LF performed the main analysis and produced figures. OA and CL identified Cluster intervals, provided the in situ dataset, and contributed to the observational spectral analysis.
PH provided the hybrid code, LF performed the numerical simulations, and together with LM, SL, AV and EP, they discussed the use and interpretation of numerical data.
All authors contributed to the global interpretation of the results, as well as to their discussion and presentation in the manuscript. All authors revised the manuscript before submission.

\section*{Funding}
This work was supported by the Programme National PNST of CNRS/INSU co-funded by CNES.
It has also been funded by Fondazione Cassa di Risparmio di Firenze through the project  HYPERCRHEL. We acknowledge PRACE for awarding us access to resource Cartesius based in the Netherlands at SURFsara through the DECI-13 (Distributed European Computing Initiative) call (project “HybTurb3D”), and CINECA for the availability of high performance computing resources and support under the ISCRA initiative (grants HP10C877C4 and HP10BUUOJM) and the program Accordo Quadro INAF-CINECA 2017-2019 (grants C4A26 and C3A22a).
L.F. was supported by Fondazione Cassa di Risparmio di Firenze, through the project “Giovani Ricercatori Protagonisti”, and by the UK Science and Technology Facilities Council (STFC) grants ST/P000622/1 and ST/T00018X/1.
PH acknowledges grant 18-08861S of the Czech Science Foundation.
OA and CL are supported by the French Centre National d'Etude Spatiales (CNES).

\section*{Acknowledgments}
Authors acknowledge useful discussions with J. Stawarz, G. Howes, and A. Pitna.

\section*{Data Availability Statement}
The in situ observations analyzed for this study can be found in the main ESA Cluster archive, \href{url}{https://csa.esac.esa.int}. 
The numerical simulations dataset analyzed for this study can be found in the Data deposit by the EUDAT infrastructure (https://www.eudat.eu/), through a persistent identifier (pid): 
\href{url}{https://b2share.eudat.eu/records/ a58135af9c9d429f92c15ce88bdfdd55}.



\begin{thebibliography}{69}
\providecommand{\natexlab}[1]{#1}
\expandafter\ifx\csname urlstyle\endcsname\relax
  \providecommand{\doi}[1]{doi:\discretionary{}{}{}#1}\else
  \providecommand{\doi}{doi:\discretionary{}{}{}\begingroup
  \urlstyle{rm}\Url}\fi
\providecommand{\selectlanguage}[1]{\relax}
\providecommand{\bibAnnoteFile}[1]{%
  \IfFileExists{#1}{\begin{quotation}\noindent\textsc{Key:} #1\\
  \textsc{Annotation:}\ \input{#1}\end{quotation}}{}}
\providecommand{\bibAnnote}[2]{%
  \begin{quotation}\noindent\textsc{Key:} #1\\
  \textsc{Annotation:}\ #2\end{quotation}}

\bibitem[{{Alexandrova} et~al.(2020){Alexandrova}, {Krishna Jagarlamudi},
  {Rossi}, {Maksimovic}, {Hellinger}, {Shprits} et~al.}]{Alexandrova_al_2020}
{Alexandrova}, O., {Krishna Jagarlamudi}, V., {Rossi}, C., {Maksimovic}, M.,
  {Hellinger}, P., {Shprits}, Y., et~al. (2020).
\newblock {Kinetic turbulence in space plasmas observed in the near-Earth and
  near-Sun solar wind}.
\newblock \emph{arXiv e-prints} , arXiv:2004.01102
\input{Alexandrova_al_2020}

\bibitem[{{Alexandrova} et~al.(2008){Alexandrova}, {Lacombe}, and
  {Mangeney}}]{Alexandrova_al_2008}
{Alexandrova}, O., {Lacombe}, C., and {Mangeney}, A. (2008).
\newblock {Spectra and anisotropy of magnetic fluctuations in the Earth's
  magnetosheath: Cluster observations}.
\newblock \emph{Annales Geophysicae} 26, 3585--3596
\input{Alexandrova_al_2008}

\bibitem[{{Alexandrova} et~al.(2012){Alexandrova}, {Lacombe}, {Mangeney},
  {Grappin}, and {Maksimovic}}]{Alexandrova_al_2012}
{Alexandrova}, O., {Lacombe}, C., {Mangeney}, A., {Grappin}, R., and
  {Maksimovic}, M. (2012).
\newblock {Solar Wind Turbulent Spectrum at Plasma Kinetic Scales}.
\newblock \emph{ApJ} 760, 121.
\newblock \doi{10.1088/0004-637X/760/2/121}
\input{Alexandrova_al_2012}

\bibitem[{{Alexandrova} et~al.(2009){Alexandrova}, {Saur}, {Lacombe},
  {Mangeney}, {Mitchell}, {Schwartz} et~al.}]{Alexandrova_al_2009}
{Alexandrova}, O., {Saur}, J., {Lacombe}, C., {Mangeney}, A., {Mitchell}, J.,
  {Schwartz}, S.~J., et~al. (2009).
\newblock {Universality of Solar-Wind Turbulent Spectrum from MHD to Electron
  Scales}.
\newblock \emph{Phys. Rev. Lett.} 103, 165003--+.
\newblock \doi{10.1103/PhysRevLett.103.165003}
\input{Alexandrova_al_2009}

\bibitem[{{Biskamp} et~al.(1996){Biskamp}, {Schwarz}, and
  {Drake}}]{Biskamp_al_1996}
{Biskamp}, D., {Schwarz}, E., and {Drake}, J.~F. (1996).
\newblock {Two-Dimensional Electron Magnetohydrodynamic Turbulence}.
\newblock \emph{Phys. Rev. Lett.} 76, 1264--1267.
\newblock \doi{10.1103/PhysRevLett.76.1264}
\input{Biskamp_al_1996}

\bibitem[{{Boldyrev} et~al.(2013){Boldyrev}, {Horaites}, {Xia}, and
  {Perez}}]{Boldyrev_al_2013}
{Boldyrev}, S., {Horaites}, K., {Xia}, Q., and {Perez}, J.~C. (2013).
\newblock {Toward a Theory of Astrophysical Plasma Turbulence at Subproton
  Scales}.
\newblock \emph{ApJ} 777, 41.
\newblock \doi{10.1088/0004-637X/777/1/41}
\input{Boldyrev_al_2013}

\bibitem[{{Boldyrev} and {Perez}(2012)}]{Boldyrev_Perez_2012}
{Boldyrev}, S. and {Perez}, J.~C. (2012).
\newblock {Spectrum of Kinetic-Alfv{\'e}n Turbulence}.
\newblock \emph{ApJ Lett.} 758, L44.
\newblock \doi{10.1088/2041-8205/758/2/L44}
\input{Boldyrev_Perez_2012}

\bibitem[{{Bruno} and {Carbone}(2013)}]{Bruno_Carbone_2013}
{Bruno}, R. and {Carbone}, V. (2013).
\newblock {The Solar Wind as a Turbulence Laboratory}.
\newblock \emph{Living Reviews in Solar Physics} 10, 2.
\newblock \doi{10.12942/lrsp-2013-2}
\input{Bruno_Carbone_2013}

\bibitem[{{Cerri} et~al.(2016){Cerri}, {Califano}, {Jenko}, {Told}, and
  {Rincon}}]{Cerri_al_2016}
{Cerri}, S.~S., {Califano}, F., {Jenko}, F., {Told}, D., and {Rincon}, F.
  (2016).
\newblock {Subproton-scale Cascades in Solar Wind Turbulence: Driven
  Hybrid-kinetic Simulations}.
\newblock \emph{ApJ Lett.} 822, L12
\input{Cerri_al_2016}

\bibitem[{{Cerri} et~al.(2019){Cerri}, {Gro\v{s}elj}, and
  {Franci}}]{Cerri_al_2019}
{Cerri}, S.~S., {Gro\v{s}elj}, D., and {Franci}, L. (2019).
\newblock {Kinetic plasma turbulence: recent insights and open questions from
  3D3V simulations}.
\newblock \emph{Frontiers in Astronomy and Space Sciences} 6, 64.
\newblock \doi{10.3389/fspas.2019.00064}
\input{Cerri_al_2019}

\bibitem[{{Cerri} et~al.(2018){Cerri}, {Kunz}, and {Califano}}]{Cerri_al_2018}
{Cerri}, S.~S., {Kunz}, M.~W., and {Califano}, F. (2018).
\newblock {Dual Phase-space Cascades in 3D Hybrid-Vlasov-Maxwell Turbulence}.
\newblock \emph{ApJ Lett.} 856, L13.
\newblock \doi{10.3847/2041-8213/aab557}
\input{Cerri_al_2018}

\bibitem[{{Cerri} et~al.(2017){Cerri}, {Servidio}, and
  {Califano}}]{Cerri_al_2017}
{Cerri}, S.~S., {Servidio}, S., and {Califano}, F. (2017).
\newblock {Kinetic Cascade in Solar-wind Turbulence: 3D3V Hybrid-kinetic
  Simulations with Electron Inertia}.
\newblock \emph{ApJ Lett.} 846, L18.
\newblock \doi{10.3847/2041-8213/aa87b0}
\input{Cerri_al_2017}

\bibitem[{{Chen}(2016)}]{Chen_2016}
{Chen}, C.~H.~K. (2016).
\newblock {Recent progress in astrophysical plasma turbulence from solar wind
  observations}.
\newblock \emph{Journal of Plasma Physics} 82, 535820602.
\newblock \doi{10.1017/S0022377816001124}
\input{Chen_2016}

\bibitem[{{Chen} et~al.(2013{\natexlab{a}}){Chen}, {Bale}, {Salem}, and
  {Maruca}}]{Chen_al_2013}
{Chen}, C.~H.~K., {Bale}, S.~D., {Salem}, C.~S., and {Maruca}, B.~A.
  (2013{\natexlab{a}}).
\newblock {Residual Energy Spectrum of Solar Wind Turbulence}.
\newblock \emph{ApJ} 770, 125.
\newblock \doi{10.1088/0004-637X/770/2/125}
\input{Chen_al_2013}

\bibitem[{{Chen} and {Boldyrev}(2017)}]{Chen_Boldyrev_2017}
{Chen}, C.~H.~K. and {Boldyrev}, S. (2017).
\newblock {Nature of Kinetic Scale Turbulence in the Earth's Magnetosheath}.
\newblock \emph{ApJ} 842, 122.
\newblock \doi{10.3847/1538-4357/aa74e0}
\input{Chen_Boldyrev_2017}

\bibitem[{{Chen} et~al.(2013{\natexlab{b}}){Chen}, {Boldyrev}, {Xia}, and
  {Perez}}]{Chen_al_2013b}
{Chen}, C.~H.~K., {Boldyrev}, S., {Xia}, Q., and {Perez}, J.~C.
  (2013{\natexlab{b}}).
\newblock {Nature of Subproton Scale Turbulence in the Solar Wind}.
\newblock \emph{Phys. Rev. Lett.} 110, 225002.
\newblock \doi{10.1103/PhysRevLett.110.225002}
\input{Chen_al_2013b}

\bibitem[{{Chen} et~al.(2010){Chen}, {Horbury}, {Schekochihin}, {Wicks},
  {Alexandrova}, and {Mitchell}}]{Chen_al_2010}
{Chen}, C.~H.~K., {Horbury}, T.~S., {Schekochihin}, A.~A., {Wicks}, R.~T.,
  {Alexandrova}, O., and {Mitchell}, J. (2010).
\newblock {Anisotropy of Solar Wind Turbulence between Ion and Electron
  Scales}.
\newblock \emph{Physical Review Letters} 104, 255002.
\newblock \doi{10.1103/PhysRevLett.104.255002}
\input{Chen_al_2010}

\bibitem[{{Chen} et~al.(2014){Chen}, {Leung}, {Boldyrev}, {Maruca}, and
  {Bale}}]{Chen_al_2014b}
{Chen}, C.~H.~K., {Leung}, L., {Boldyrev}, S., {Maruca}, B.~A., and {Bale},
  S.~D. (2014).
\newblock {Ion-scale spectral break of solar wind turbulence at high and low
  beta}.
\newblock \emph{Geophys. Res. Let.} 41, 8081--8088.
\newblock \doi{10.1002/2014GL062009}
\input{Chen_al_2014b}

\bibitem[{{Chen} et~al.(2012{\natexlab{a}}){Chen}, {Mallet}, {Schekochihin},
  {Horbury}, {Wicks}, and {Bale}}]{Chen_al_2012b}
{Chen}, C.~H.~K., {Mallet}, A., {Schekochihin}, A.~A., {Horbury}, T.~S.,
  {Wicks}, R.~T., and {Bale}, S.~D. (2012{\natexlab{a}}).
\newblock {Three-dimensional Structure of Solar Wind Turbulence}.
\newblock \emph{ApJ} 758, 120.
\newblock \doi{10.1088/0004-637X/758/2/120}
\input{Chen_al_2012b}

\bibitem[{{Chen} et~al.(2012{\natexlab{b}}){Chen}, {Salem}, {Bonnell}, {Mozer},
  and {Bale}}]{Chen_al_2012}
{Chen}, C.~H.~K., {Salem}, C.~S., {Bonnell}, J.~W., {Mozer}, F.~S., and {Bale},
  S.~D. (2012{\natexlab{b}}).
\newblock {Density Fluctuation Spectrum of Solar Wind Turbulence between Ion
  and Electron Scales}.
\newblock \emph{Phys. Rev. Lett.} 109, 035001
\input{Chen_al_2012}

\bibitem[{Dasso et~al.(2005)Dasso, Milano, Matthaeus, and
  Smith}]{Dasso_al_2005}
Dasso, S., Milano, L.~J., Matthaeus, W.~H., and Smith, C.~W. (2005).
\newblock Anisotropy in fast and slow solar wind fluctuations.
\newblock \emph{ApJ} 635, L181--L184.
\newblock \doi{10.1086/499559}
\input{Dasso_al_2005}

\bibitem[{{Franci} et~al.(2018{\natexlab{a}}){Franci}, {Hellinger}, {Guarrasi},
  {Chen}, {Papini}, {Verdini} et~al.}]{Franci_al_2018b}
{Franci}, L., {Hellinger}, P., {Guarrasi}, M., {Chen}, C.~H.~K., {Papini}, E.,
  {Verdini}, A., et~al. (2018{\natexlab{a}}).
\newblock {Three-dimensional simulations of solar wind turbulence with the
  hybrid code CAMELIA}.
\newblock In \emph{Journal of Physics Conference Series}. vol. 1031 of
  \emph{Journal of Physics Conference Series}, 012002.
\newblock \doi{10.1088/1742-6596/1031/1/012002}
\input{Franci_al_2018b}

\bibitem[{{Franci} et~al.(2015{\natexlab{a}}){Franci}, {Landi}, {Matteini},
  {Verdini}, and {Hellinger}}]{Franci_al_2015b}
{Franci}, L., {Landi}, S., {Matteini}, L., {Verdini}, A., and {Hellinger}, P.
  (2015{\natexlab{a}}).
\newblock {High-resolution Hybrid Simulations of Kinetic Plasma Turbulence at
  Proton Scales}.
\newblock \emph{ApJ} 812, 21.
\newblock \doi{10.1088/0004-637X/812/1/21}
\input{Franci_al_2015b}

\bibitem[{{Franci} et~al.(2016){Franci}, {Landi}, {Matteini}, {Verdini}, and
  {Hellinger}}]{Franci_al_2016}
{Franci}, L., {Landi}, S., {Matteini}, L., {Verdini}, A., and {Hellinger}, P.
  (2016).
\newblock {Plasma Beta Dependence of the Ion-scale Spectral Break of Solar Wind
  Turbulence: High-resolution 2D Hybrid Simulations}.
\newblock \emph{ApJ} 833, 91.
\newblock \doi{10.3847/1538-4357/833/1/91}
\input{Franci_al_2016}

\bibitem[{{Franci} et~al.(2018{\natexlab{b}}){Franci}, {Landi}, {Verdini},
  {Matteini}, and {Hellinger}}]{Franci_al_2018}
{Franci}, L., {Landi}, S., {Verdini}, A., {Matteini}, L., and {Hellinger}, P.
  (2018{\natexlab{b}}).
\newblock {Solar Wind Turbulent Cascade from MHD to Sub-ion Scales: Large-size
  3D Hybrid Particle-in-cell Simulations}.
\newblock \emph{ApJ} 853, 26.
\newblock \doi{10.3847/1538-4357/aaa3e8}
\input{Franci_al_2018}

\bibitem[{{Franci} et~al.(2015{\natexlab{b}}){Franci}, {Verdini}, {Matteini},
  {Landi}, and {Hellinger}}]{Franci_al_2015a}
{Franci}, L., {Verdini}, A., {Matteini}, L., {Landi}, S., and {Hellinger}, P.
  (2015{\natexlab{b}}).
\newblock {Solar Wind Turbulence from MHD to Sub-ion Scales: High-resolution
  Hybrid Simulations}.
\newblock \emph{ApJ Lett.} 804, L39.
\newblock \doi{10.1088/2041-8205/804/2/L39}
\input{Franci_al_2015a}

\bibitem[{{Gro\v{s}elj} et~al.(2017){Gro\v{s}elj}, {Cerri}, {Ba{\~n}{\'o}n
  Navarro}, {Willmott}, {Told}, {Loureiro} et~al.}]{Groselj_al_2017}
{Gro\v{s}elj}, D., {Cerri}, S.~S., {Ba{\~n}{\'o}n Navarro}, A., {Willmott}, C.,
  {Told}, D., {Loureiro}, N.~F., et~al. (2017).
\newblock {Fully Kinetic versus Reduced-kinetic Modeling of Collisionless
  Plasma Turbulence}.
\newblock \emph{ApJ} 847, 28.
\newblock \doi{10.3847/1538-4357/aa894d}
\input{Groselj_al_2017}

\bibitem[{{Gro{\v{s}}elj} et~al.(2019){Gro{\v{s}}elj}, {Chen}, {Mallet},
  {Samtaney}, {Schneider}, and {Jenko}}]{Groselj_al_2019}
{Gro{\v{s}}elj}, D., {Chen}, C. H.~K., {Mallet}, A., {Samtaney}, R.,
  {Schneider}, K., and {Jenko}, F. (2019).
\newblock {Kinetic Turbulence in Astrophysical Plasmas: Waves and/or
  Structures?}
\newblock \emph{Physical Review X} 9, 031037.
\newblock \doi{10.1103/PhysRevX.9.031037}
\input{Groselj_al_2019}

\bibitem[{{Hamilton} et~al.(2008){Hamilton}, {Smith}, {Vasquez}, and
  {Leamon}}]{Hamilton_al_2008}
{Hamilton}, K., {Smith}, C.~W., {Vasquez}, B.~J., and {Leamon}, R.~J. (2008).
\newblock {Anisotropies and helicities in the solar wind inertial and
  dissipation ranges at 1 AU}.
\newblock \emph{Journal of Geophysical Research (Space Physics)} 113, A01106.
\newblock \doi{10.1029/2007JA012559}
\input{Hamilton_al_2008}

\bibitem[{Hellinger et~al.(2018)Hellinger, Verdini, Landi, Franci, and
  Matteini}]{Hellinger_al_2018}
Hellinger, P., Verdini, A., Landi, S., Franci, L., and Matteini, L. (2018).
\newblock von {K\'arm\'an-Howarth} equation for {Hall} magnetohydrodynamics:
  Hybrid simulations.
\newblock \emph{Astrophys. J. Lett.} 857, L19.
\newblock \doi{10.3847/2041-8213/aabc06}
\input{Hellinger_al_2018}

\bibitem[{{Horbury} and {Balogh}(2001)}]{Horbury_Balogh_2001}
{Horbury}, T.~S. and {Balogh}, A. (2001).
\newblock {Evolution of magnetic field fluctuations in high-speed solar wind
  streams: Ulysses and Helios observations}.
\newblock \emph{J. Geophys. Res.} 106, 15929--15940.
\newblock \doi{10.1029/2000JA000108}
\input{Horbury_Balogh_2001}

\bibitem[{{Horbury} et~al.(2008){Horbury}, {Forman}, and
  {Oughton}}]{Horbury_al_2008}
{Horbury}, T.~S., {Forman}, M., and {Oughton}, S. (2008).
\newblock {Anisotropic Scaling of Magnetohydrodynamic Turbulence}.
\newblock \emph{Phys. Rev. Lett.} 101, 175005--+.
\newblock \doi{10.1103/PhysRevLett.101.175005}
\input{Horbury_al_2008}

\bibitem[{{Howes} et~al.(2008){Howes}, {Dorland}, {Cowley}, {Hammett},
  {Quataert}, {Schekochihin} et~al.}]{Howes_al_2008}
{Howes}, G.~G., {Dorland}, W., {Cowley}, S.~C., {Hammett}, G.~W., {Quataert},
  E., {Schekochihin}, A.~A., et~al. (2008).
\newblock {Kinetic Simulations of Magnetized Turbulence in Astrophysical
  Plasmas}.
\newblock \emph{Phys. Rev. Lett.} 100, 065004.
\newblock \doi{10.1103/PhysRevLett.100.065004}
\input{Howes_al_2008}

\bibitem[{{Howes} et~al.(2011){Howes}, {Tenbarge}, {Dorland}, {Quataert},
  {Schekochihin}, {Numata} et~al.}]{Howes_al_2011}
{Howes}, G.~G., {Tenbarge}, J.~M., {Dorland}, W., {Quataert}, E.,
  {Schekochihin}, A.~A., {Numata}, R., et~al. (2011).
\newblock {Gyrokinetic Simulations of Solar Wind Turbulence from Ion to
  Electron Scales}.
\newblock \emph{Physical Review Letters} 107, 035004.
\newblock \doi{10.1103/PhysRevLett.107.035004}
\input{Howes_al_2011}

\bibitem[{{Jovanovic} et~al.(2020){Jovanovic}, {Alexandrova}, {Maksimovic}, and
  {Belic}}]{Jovanovic_al_2020}
{Jovanovic}, D., {Alexandrova}, O., {Maksimovic}, M., and {Belic}, M. (2020).
\newblock {Fluid theory of coherent magnetic vortices in high-beta space
  plasmas}.
\newblock \emph{apj} , arXiv:1705.02913
\input{Jovanovic_al_2020}

\bibitem[{{Kiyani} et~al.(2009){Kiyani}, {Chapman}, {Khotyaintsev}, {Dunlop},
  and {Sahraoui}}]{Kiyani_al_2009}
{Kiyani}, K.~H., {Chapman}, S.~C., {Khotyaintsev}, Y.~V., {Dunlop}, M.~W., and
  {Sahraoui}, F. (2009).
\newblock {Global Scale-Invariant Dissipation in Collisionless Plasma
  Turbulence}.
\newblock \emph{prl} 103, 075006.
\newblock \doi{10.1103/PhysRevLett.103.075006}
\input{Kiyani_al_2009}

\bibitem[{{Kiyani} et~al.(2013){Kiyani}, {Chapman}, {Sahraoui}, {Hnat},
  {Fauvarque}, and {Khotyaintsev}}]{Kiyani_al_2013}
{Kiyani}, K.~H., {Chapman}, S.~C., {Sahraoui}, F., {Hnat}, B., {Fauvarque}, O.,
  and {Khotyaintsev}, Y.~V. (2013).
\newblock {Enhanced Magnetic Compressibility and Isotropic Scale Invariance at
  Sub-ion Larmor Scales in Solar Wind Turbulence}.
\newblock \emph{ApJ} 763, 10.
\newblock \doi{10.1088/0004-637X/763/1/10}
\input{Kiyani_al_2013}

\bibitem[{{Lacombe} et~al.(2017){Lacombe}, {Alexandrova}, and
  {Matteini}}]{Lacombe_al_2017}
{Lacombe}, C., {Alexandrova}, O., and {Matteini}, L. (2017).
\newblock {Anisotropies of the Magnetic Field Fluctuations at Kinetic Scales in
  the Solar Wind: Cluster Observations}.
\newblock \emph{ApJ} 848, 45.
\newblock \doi{10.3847/1538-4357/aa8c06}
\input{Lacombe_al_2017}

\bibitem[{{Landi} et~al.(2019){Landi}, {Franci}, {Papini}, {Verdini},
  {Matteini}, and {Hellinger}}]{Landi_al_2019}
{Landi}, S., {Franci}, L., {Papini}, E., {Verdini}, A., {Matteini}, L., and
  {Hellinger}, P. (2019).
\newblock {Spectral anisotropies and intermittency of plasma turbulence at ion
  kinetic scales}.
\newblock \emph{arXiv e-prints} , arXiv:1904.03903
\input{Landi_al_2019}

\bibitem[{{Loureiro} and {Boldyrev}(2017)}]{Loureiro_Boldyrev_2017}
{Loureiro}, N.~F. and {Boldyrev}, S. (2017).
\newblock {Collisionless Reconnection in Magnetohydrodynamic and Kinetic
  Turbulence}.
\newblock \emph{ApJ} 850, 182.
\newblock \doi{10.3847/1538-4357/aa9754}
\input{Loureiro_Boldyrev_2017}

\bibitem[{{Mallet} et~al.(2017){Mallet}, {Schekochihin}, and
  {Chandran}}]{Mallet_al_2017}
{Mallet}, A., {Schekochihin}, A.~A., and {Chandran}, B.~D.~G. (2017).
\newblock {Disruption of sheet-like structures in Alfv{\'e}nic turbulence by
  magnetic reconnection} 468, 4862--4871.
\newblock \doi{10.1093/mnras/stx670}
\input{Mallet_al_2017}

\bibitem[{{Matteini} et~al.(2017){Matteini}, {Alexandrova}, {Chen}, and
  {Lacombe}}]{Matteini_al_2017}
{Matteini}, L., {Alexandrova}, O., {Chen}, C.~H.~K., and {Lacombe}, C. (2017).
\newblock {Electric and magnetic spectra from MHD to electron scales in the
  magnetosheath}.
\newblock \emph{MNRAS} 466, 945--951.
\newblock \doi{10.1093/mnras/stw3163}
\input{Matteini_al_2017}

\bibitem[{{Matthaeus} et~al.(1990){Matthaeus}, {Goldstein}, and
  {Roberts}}]{Matthaeus_al_1990}
{Matthaeus}, W.~H., {Goldstein}, M.~L., and {Roberts}, D.~A. (1990).
\newblock {Evidence for the presence of quasi-two-dimensional nearly
  incompressible fluctuations in the solar wind}.
\newblock \emph{J. Geophys. Res.} 95, 20673--20683.
\newblock \doi{10.1029/JA095iA12p20673}
\input{Matthaeus_al_1990}

\bibitem[{{Matthews}(1994)}]{Matthews_1994}
{Matthews}, A.~P. (1994).
\newblock Current advance method and cyclic leapfrog for 2d multispecies hybrid
  plasma simulations.
\newblock \emph{J. Comp. Phys.} 112, 102--116
\input{Matthews_1994}

\bibitem[{Osman and Horbury(2006)}]{Osman_Horbury_2006}
Osman, K.~T. and Horbury, T.~S. (2006).
\newblock Multispacecraft measurement of anisotropic correlation functions in
  solar wind turbulence.
\newblock \emph{ApJ} 654, L103--L106.
\newblock \doi{10.1086/510906}
\input{Osman_Horbury_2006}

\bibitem[{{Papini} et~al.(2019){Papini}, {Franci}, {Landi}, {Verdini},
  {Matteini}, and {Hellinger}}]{Papini_al_2018}
{Papini}, E., {Franci}, L., {Landi}, S., {Verdini}, A., {Matteini}, L., and
  {Hellinger}, P. (2019).
\newblock {Can Hall Magnetohydrodynamics Explain Plasma Turbulence at Sub-ion
  Scales?}
\newblock \emph{apj} 870, 52.
\newblock \doi{10.3847/1538-4357/aaf003}
\input{Papini_al_2018}

\bibitem[{{Passot} and {Sulem}(2015)}]{Passot_Sulem_2015}
{Passot}, T. and {Sulem}, P.~L. (2015).
\newblock {A Model for the Non-universal Power Law of the Solar Wind
  Sub-ion-scale Magnetic Spectrum}.
\newblock \emph{ApJ Lett.} 812, L37.
\newblock \doi{10.1088/2041-8205/812/2/L37}
\input{Passot_Sulem_2015}

\bibitem[{{Passot} et~al.(2017){Passot}, {Sulem}, and {Tassi}}]{Passot_al_2017}
{Passot}, T., {Sulem}, P.~L., and {Tassi}, E. (2017).
\newblock {Electron-scale reduced fluid models with gyroviscous effects}.
\newblock \emph{Journal of Plasma Physics} 83, 715830402.
\newblock \doi{10.1017/S0022377817000514}
\input{Passot_al_2017}

\bibitem[{{Perrone} et~al.(2017){Perrone}, {Alexandrova}, {Roberts}, {Lion},
  {Lacombe}, {Walsh} et~al.}]{Perrone_al_2017}
{Perrone}, D., {Alexandrova}, O., {Roberts}, O.~W., {Lion}, S., {Lacombe}, C.,
  {Walsh}, A., et~al. (2017).
\newblock {Coherent Structures at Ion Scales in Fast Solar Wind: Cluster
  Observations}.
\newblock \emph{apj} 849, 49.
\newblock \doi{10.3847/1538-4357/aa9022}
\input{Perrone_al_2017}

\bibitem[{{Pit{\v{n}}a} et~al.(2019){Pit{\v{n}}a}, {{\v{S}}afr{\'a}nkov{\'a}},
  {N{\v{e}}me{\v{c}}ek}, {Franci}, {Pi}, and {Montagud Camps}}]{Pitna_al_2019}
{Pit{\v{n}}a}, A., {{\v{S}}afr{\'a}nkov{\'a}}, J., {N{\v{e}}me{\v{c}}ek}, Z.,
  {Franci}, L., {Pi}, G., and {Montagud Camps}, V. (2019).
\newblock {Characteristics of Solar Wind Fluctuations at and below Ion Scales}.
\newblock \emph{ApJ} 879, 82.
\newblock \doi{10.3847/1538-4357/ab22b8}
\input{Pitna_al_2019}

\bibitem[{{Podesta} and {TenBarge}(2012)}]{Podesta_TenBarge_2012}
{Podesta}, J.~J. and {TenBarge}, J.~M. (2012).
\newblock {Scale dependence of the variance anisotropy near the proton
  gyroradius scale: Additional evidence for kinetic Alfv{\'e}n waves in the
  solar wind at 1 AU}.
\newblock \emph{Journal of Geophysical Research (Space Physics)} 117, A10106.
\newblock \doi{10.1029/2012JA017724}
\input{Podesta_TenBarge_2012}

\bibitem[{{Roberts} et~al.(2017){Roberts}, {Narita}, and
  {Escoubet}}]{Roberts_al_2017}
{Roberts}, O.~W., {Narita}, Y., and {Escoubet}, C.~P. (2017).
\newblock {Direct Measurement of Anisotropic and Asymmetric Wave Vector
  Spectrum in Ion-scale Solar Wind Turbulence}.
\newblock \emph{ApJ Lett.} 851, L11.
\newblock \doi{10.3847/2041-8213/aa9bf3}
\input{Roberts_al_2017}

\bibitem[{{Sahraoui} et~al.(2010){Sahraoui}, {Goldstein}, {Belmont}, {Canu},
  and {Rezeau}}]{Sahraoui_al_2010}
{Sahraoui}, F., {Goldstein}, M.~L., {Belmont}, G., {Canu}, P., and {Rezeau}, L.
  (2010).
\newblock {Three Dimensional Anisotropic k Spectra of Turbulence at Subproton
  Scales in the Solar Wind}.
\newblock \emph{Phys. Rev. Lett.} 105, 131101.
\newblock \doi{10.1103/PhysRevLett.105.131101}
\input{Sahraoui_al_2010}

\bibitem[{{Sahraoui} et~al.(2013){Sahraoui}, {Huang}, {Belmont}, {Goldstein},
  {R{\'e}tino}, {Robert} et~al.}]{Sahraoui_al_2013}
{Sahraoui}, F., {Huang}, S.~Y., {Belmont}, G., {Goldstein}, M.~L.,
  {R{\'e}tino}, A., {Robert}, P., et~al. (2013).
\newblock {Scaling of the Electron Dissipation Range of Solar Wind Turbulence}.
\newblock \emph{ApJ} 777, 15.
\newblock \doi{10.1088/0004-637X/777/1/15}
\input{Sahraoui_al_2013}

\bibitem[{{Salem} et~al.(2012){Salem}, {Howes}, {Sundkvist}, {Bale}, {Chaston},
  {Chen} et~al.}]{Salem_al_2012}
{Salem}, C.~S., {Howes}, G.~G., {Sundkvist}, D., {Bale}, S.~D., {Chaston},
  C.~C., {Chen}, C.~H.~K., et~al. (2012).
\newblock {Identification of Kinetic Alfv{\'e}n Wave Turbulence in the Solar
  Wind}.
\newblock \emph{ApJ Lett.} 745, L9.
\newblock \doi{10.1088/2041-8205/745/1/L9}
\input{Salem_al_2012}

\bibitem[{{Saur} and {Bieber}(1999)}]{Saur_Bieber_1999}
{Saur}, J. and {Bieber}, J.~W. (1999).
\newblock {Geometry of low-frequency solar wind magnetic turbulence: Evidence
  for radially aligned Alf{\'e}nic fluctuations}.
\newblock \emph{J. Geophys. Res.} 104, 9975--9988.
\newblock \doi{10.1029/1998JA900077}
\input{Saur_Bieber_1999}

\bibitem[{{Schekochihin} et~al.(2009){Schekochihin}, {Cowley}, {Dorland},
  {Hammett}, {Howes}, {Quataert} et~al.}]{Schekochihin_al_2009}
{Schekochihin}, A.~A., {Cowley}, S.~C., {Dorland}, W., {Hammett}, G.~W.,
  {Howes}, G.~G., {Quataert}, E., et~al. (2009).
\newblock {Astrophysical Gyrokinetics: Kinetic and Fluid Turbulent Cascades in
  Magnetized Weakly Collisional Plasmas}.
\newblock \emph{Astrophys. J. Suppl.} 182, 310--377.
\newblock \doi{10.1088/0067-0049/182/1/310}
\input{Schekochihin_al_2009}

\bibitem[{{Schreiner} and {Saur}(2017)}]{Schreiner_Saur_2017}
{Schreiner}, A. and {Saur}, J. (2017).
\newblock {A Model for Dissipation of Solar Wind Magnetic Turbulence by Kinetic
  Alfv{\'e}n Waves at Electron Scales: Comparison with Observations}.
\newblock \emph{apj} 835, 133.
\newblock \doi{10.3847/1538-4357/835/2/133}
\input{Schreiner_Saur_2017}

\bibitem[{{Smith} et~al.(2006){Smith}, {Vasquez}, and
  {Hamilton}}]{Smith_al_2006b}
{Smith}, C.~W., {Vasquez}, B.~J., and {Hamilton}, K. (2006).
\newblock {Interplanetary magnetic fluctuation anisotropy in the inertial
  range}.
\newblock \emph{Journal of Geophysical Research (Space Physics)} 111, A09111.
\newblock \doi{10.1029/2006JA011651}
\input{Smith_al_2006b}

\bibitem[{Stawarz et~al.(2016)Stawarz, Eriksson, Wilder, Ergun, Schwartz,
  Pouquet et~al.}]{Stawarz_al_2016}
Stawarz, J.~E., Eriksson, S., Wilder, F.~D., Ergun, R.~E., Schwartz, S.~J.,
  Pouquet, A., et~al. (2016).
\newblock Observations of turbulence in a kelvin-helmholtz event on 8 september
  2015 by the magnetospheric multiscale mission.
\newblock \emph{J. Geophys. Res.} \doi{10.1002/2016JA023458}.
\newblock 2016JA023458
\input{Stawarz_al_2016}

\bibitem[{{Sulem} et~al.(2016){Sulem}, {Passot}, {Laveder}, and
  {Borgogno}}]{sulem_al_2016}
{Sulem}, P.~L., {Passot}, T., {Laveder}, D., and {Borgogno}, D. (2016).
\newblock {Influence of the Nonlinearity Parameter on the Solar Wind Sub-ion
  Magnetic Energy Spectrum: FLR-Landau Fluid Simulations}.
\newblock \emph{ApJ} 818, 66.
\newblock \doi{10.3847/0004-637X/818/1/66}
\input{sulem_al_2016}

\bibitem[{{TenBarge} et~al.(2012){TenBarge}, {Podesta}, {Klein}, and
  {Howes}}]{TenBarge_al_2012}
{TenBarge}, J.~M., {Podesta}, J.~J., {Klein}, K.~G., and {Howes}, G.~G. (2012).
\newblock {Interpreting Magnetic Variance Anisotropy Measurements in the Solar
  Wind}.
\newblock \emph{apj} 753, 107.
\newblock \doi{10.1088/0004-637X/753/2/107}
\input{TenBarge_al_2012}

\bibitem[{{Turner} et~al.(2011){Turner}, {Gogoberidze}, {Chapman}, {Hnat}, and
  {M{\"u}ller}}]{Turner_al_2011}
{Turner}, A.~J., {Gogoberidze}, G., {Chapman}, S.~C., {Hnat}, B., and
  {M{\"u}ller}, W.~C. (2011).
\newblock {Nonaxisymmetric Anisotropy of Solar Wind Turbulence}.
\newblock \emph{Phys. Rev. Lett.} 107, 095002.
\newblock \doi{10.1103/PhysRevLett.107.095002}
\input{Turner_al_2011}

\bibitem[{{{\v S}afr{\'a}nkov{\'a}} et~al.(2013){{\v S}afr{\'a}nkov{\'a}},
  {N{\v e}me{\v c}ek}, {P{\v r}ech}, and {Zastenker}}]{Safrankova_al_2013}
{{\v S}afr{\'a}nkov{\'a}}, J., {N{\v e}me{\v c}ek}, Z., {P{\v r}ech}, L., and
  {Zastenker}, G.~N. (2013).
\newblock {Ion Kinetic Scale in the Solar Wind Observed}.
\newblock \emph{Phys. Rev. Lett.} 110, 025004.
\newblock \doi{10.1103/PhysRevLett.110.025004}
\input{Safrankova_al_2013}

\bibitem[{{Verdini} and {Grappin}(2015)}]{Verdini_Grappin_2015}
{Verdini}, A. and {Grappin}, R. (2015).
\newblock {Imprints of Expansion on the Local Anisotropy of Solar Wind
  Turbulence}.
\newblock \emph{ApJ Lett.} 808, L34.
\newblock \doi{10.1088/2041-8205/808/2/L34}
\input{Verdini_Grappin_2015}

\bibitem[{{Verdini} et~al.(2019){Verdini}, {Grappin}, {Alexandrova}, {Franci},
  {Landi}, {Matteini} et~al.}]{Verdini_al_2019}
{Verdini}, A., {Grappin}, R., {Alexandrova}, O., {Franci}, L., {Landi}, S.,
  {Matteini}, L., et~al. (2019).
\newblock {Three-dimensional local anisotropy of velocity fluctuations in the
  solar wind}.
\newblock \emph{MNRAS} 486, 3006--3018.
\newblock \doi{10.1093/mnras/stz1041}
\input{Verdini_al_2019}

\bibitem[{{Verdini} et~al.(2018){Verdini}, {Grappin}, {Alexandrova}, and
  {Lion}}]{Verdini_al_2018}
{Verdini}, A., {Grappin}, R., {Alexandrova}, O., and {Lion}, S. (2018).
\newblock {3D Anisotropy of Solar Wind Turbulence, Tubes, or Ribbons?}
\newblock \emph{ApJ} 853, 85.
\newblock \doi{10.3847/1538-4357/aaa433}
\input{Verdini_al_2018}

\bibitem[{{Wang} et~al.(2020){Wang}, {He}, {Alexandrova}, {Dunlop}, and
  {Perrone}}]{Wang_al_2020}
{Wang}, T., {He}, J., {Alexandrova}, O., {Dunlop}, M., and {Perrone}, D.
  (2020).
\newblock {Observational Quantification of Three-dimensional Anisotropies and
  Scalings of Space Plasma Turbulence at Kinetic Scales}.
\newblock \emph{ApJ} 898, 91.
\newblock \doi{10.3847/1538-4357/ab99ca}
\input{Wang_al_2020}

\bibitem[{{Wicks} et~al.(2010){Wicks}, {Horbury}, {Chen}, and
  {Schekochihin}}]{Wicks_al_2010}
{Wicks}, R.~T., {Horbury}, T.~S., {Chen}, C.~H.~K., and {Schekochihin}, A.~A.
  (2010).
\newblock {Power and spectral index anisotropy of the entire inertial range of
  turbulence in the fast solar wind}.
\newblock \emph{MNRAS} 407, L31--L35.
\newblock \doi{10.1111/j.1745-3933.2010.00898.x}
\input{Wicks_al_2010}

\end{thebibliography}

\end{document}